\documentclass[prb,showpacs,twocolumn,aps,superscriptaddress,a4paper]{revtex4-1}
\usepackage{dcolumn,amssymb,amsmath,amsfonts,graphicx,latexsym,color,braket}

\begin{document}

\title{Universal nonanalytic behavior of the Hall conductance
in a Chern insulator at the topologically driven nonequilibrium phase transition}

\author{Pei Wang}
\email{wangpei@zjut.edu.cn}
\affiliation{Department of Physics, Zhejiang Normal University, Jinhua 321004, China}
\affiliation{Institute for Theoretical Physics, Georg-August-Universit\"{a}t G\"{o}ttingen,
Friedrich-Hund-Platz 1, G\"{o}ttingen 37077, Germany}

\author{Markus Schmitt}
\affiliation{Institute for Theoretical Physics, Georg-August-Universit\"{a}t G\"{o}ttingen,
Friedrich-Hund-Platz 1, G\"{o}ttingen 37077, Germany}

\author{Stefan Kehrein}
\affiliation{Institute for Theoretical Physics, Georg-August-Universit\"{a}t G\"{o}ttingen,
Friedrich-Hund-Platz 1, G\"{o}ttingen 37077, Germany}

\date{\today}

\begin{abstract}
We study the Hall conductance of a Chern insulator after a global quench
of the Hamiltonian.
The Hall conductance in the long time limit is obtained by
applying the linear response theory to the diagonal ensemble.
It is expressed as the integral of the Berry curvature weighted by
the occupation number over the Brillouin zone.
We identify a topologically driven nonequilibrium phase transition,
which is indicated by the nonanalyticity of the Hall conductance as a function
of the energy gap $m_f$ in the post-quench Hamiltonian $\hat H_f$.
The topological invariant for the quenched state is
the winding number of the Green's function $W$, which equals the
Chern number for the ground state of $\hat H_f$. In the limit $m_f\to 0$,
the derivative of the Hall conductance with respect to $m_f$
is proportional to $\ln |m_f|$, with the constant of proportionality being
the ratio of the change of $W$ at $m_f=0$ to the energy
gap in the initial state.
This nonanalytic behavior is universal in two-band Chern insulators such as
the Dirac model, the Haldane model, or the Kitaev honeycomb model
in the fermionic basis.
\end{abstract}

\maketitle

\section{Introduction}
The notions of topological order and topological invariant were
introduced into condensed matter physics
for classifying certain states of matter
that cannot be classified by broken symmetries.
Their change in the ground state of a system
is accompanied by a topological phase transition.
It is well known that the topological order or
the topological invariant are robust against
local perturbations to the Hamiltonian.
But it is not clear whether they are
also robust against a global quench of the Hamiltonian,
and what is the proper way of defining them
in a quenched state far from equilibrium.
Recently, these questions drew attention~\cite{tsomokos09,
halasz13,rahmani10,foster,dong14,fosterprb,wang14,perfetto13,
chung14,alessiol,caio,wang15,bermudez09,bermudez10,degottardi11,patela13,
rajak14a,rajak14b,sacramento14} due to their relevance with
the implementation of topological quantum computing.

Suppose that the system is initially in the ground state
of the Hamiltonian $\hat H_i$. At the time $t=0$,
the Hamiltonian is suddenly changed from $\hat H_i$ to $\hat H_f$. The system
is then driven out of equilibrium, and the wave function evolves unitarily.
In the toric code model, the topological entropy of
the unitarily-evolving wave function is found to keep
invariant~\cite{tsomokos09,halasz13,rahmani10} after a quench.
In general, the long-range entanglement in a wave function
cannot be changed by local unitary transformations~\cite{chen10}. Therefore,
the topological entropy of the unitarily-evolving wave function
keeps a constant if $\hat H_f$ has no long-range interaction.
Similarly, the Chern number of the unitarily-evolving wave function in a Chern
insulator~\cite{alessiol,caio,wang15} is found to be independent of $\hat H_f$.
If one chose it as the topological order parameter of the
quenched state, the topological order would be always robust
against a quench of the Hamiltonian.

However, in the p-wave superfluid or the s-wave superfluid with spin-orbit
coupling, the winding number of the Green's function depends on both $\hat H_i$
and $\hat H_f$~\cite{foster,fosterprb,dong14,wang14}.
And in the topological superconductor with proximity-induced superconductivity,
the topological properties of the quenched state were argued
to be $\hat H_f$-dependent~\cite{perfetto13,chung14,sacramento14}.
When the initial state is in a topologically nontrivial phase,
the quenched state in the long time limit is
in the trivial (nontrivial) phase
if the ground state of $\hat H_f$ is in the trivial (nontrivial) phase.
This conclusion is supported by
the study of the Majorana order parameter\cite{perfetto13},
the entanglement spectrum~\cite{chung14} and the dynamics of edge states~\cite{sacramento14}.
Especially, in the thermodynamic limit, the survival probability of the Majorana edge modes
decays to a finite value if the quench is within the same topological
phase. But it decays to zero if the quench is across the phase boundary~\cite{sacramento14}.

Up to now, the definition of the topological order parameter in a quenched state
is ambiguous. Different topological invariants which are equivalent
in a ground state might be dramatically different in a quenched state.
This is clearly demonstrated in the p-wave superfluid~\cite{fosterprb}, in which
the winding number of the Anderson pseudo spin texture
is independent of $\hat H_f$ but that of the Green's function is $\hat H_f$-dependent.
It is then necessary to study which topological invariant is
experimentally relevant.

The order parameter is defined to distinguish different
phases at a phase transition. A phase transition can be indicated
in an experimentally-relevant way by the nonanalyticity
of the observables.
And the topological invariant was introduced to explain
phase transitions that are beyond the conventional framework of symmetry breaking.
For example, the Chern number~\cite{TKNN} was introduced in the
quantum Hall effect~\cite{klitzing} to explain the jump of the Hall conductance
at some magnetic fields.
Following this logic, we study
the phase transitions in the quenched states,
which are indicated by the non\-analyticity
of a measurable observable as a function of the parameters of
the quenched state, i.e., the parameters in $\hat H_i$ or $\hat H_f$.
The experimentally-relevant topological invariant for the quenched state
is defined in such a way that its change accompanies the phase transition.

A topologically driven phase transition in the quenched state
has been argued to exist in the one-dimensional superfluid with spin-orbit coupling,
in which the superfluid order parameter and the tunneling conductance at the edge
were calculated numerically~\cite{wang14}.
However, unambiguous evidence for the nonanalyticity
of observables cannot be obtained from the numerics.
In this paper, we study the quenched state of a Chern insulator.
We argue that the Hall conductance in the long time limit after the quench
can be expressed as the integral of the Berry curvature
weighted by the occupation number over the Brillouin zone.
In a generic two-band Chern insulator, the Hall conductance
is not quantized, but is a continuous function of the energy gap $m_f$ in the
post-quench Hamiltonian $\hat H_f$. However, the derivative
of the Hall conductance with respect to $m_f$ is logarithmically
divergent in the limit $m_f \to 0$ if the Chern number for the ground state
of $\hat H_f$ changes at $m_f=0$. And the prefactor of the
logarithm is the ratio of the change of the Chern number
to the energy gap in the initial state. We strictly prove this
statement by relating the nonanalyticity of the Hall conductance to the spectrum of $\hat H_f$
nearby the gap closing point in the Brillouin zone.
The experimentally-relevant topological
invariant for the quenched state is the winding number of
the Green's function, which is equal to the Chern number
for the ground state of $\hat H_f$ but is generally different from
the Chern number of the unitarily-evolving wave function.
We then identify a nonequilibrium
phase transition in the quenched state, which
has a topological nature in the sense that the nonanalyticity
is determined by the change of the topological invariant at the transition.
The nonanalytic behavior of the Hall conductance is universal,
being independent of the symmetry of the model or
local perturbations to $\hat H_i$ and $\hat H_f$.
Our findings serve as a benchmark
in the future study of the topologically driven phase transitions in the quenched states.

The contents of the paper are arranged as follows.
We derive the formula for the quench-state Hall conductance in a
$N$-band Chern insulator in Sec.~\ref{sec:hallgeneral}, and
show its form in a two-band Chern insulator in Sec.~\ref{sec:halltwoband}.
In Sec.~\ref{sec:nonanalytic}, we apply our formalism to
three models, namely the Dirac model, the Haldane model,
and the Kitaev honeycomb model. We discuss how to address
the nonanalytic behavior of the Hall conductance.
In Sec.~\ref{sec:generalnonanalytic}, we extend to a general
two-band Chern insulator and prove the universal nonanalytic behavior
of the Hall conductance.
We discuss the topological invariant for the quenched state
in Sec.~\ref{sec:topologicalinvariant}.
At last, a concluding section summarizes our results.

\section{Hall conductance of the quenched state}
\label{sec:hallgeneral}

Let us consider a $N$-component Fermi gas in two dimensions.
Its Hamiltonian in momentum space is written as
\begin{equation}\label{eq:orginaloperatorH}
 \hat H= \sum_{\vec{k}} \hat c^\dag_{\vec{k}} \mathcal{H}_{\vec{k}} \hat c_{\vec{k}},
\end{equation}
where $\hat c_{\vec{k}} = \left(\hat c_{\vec{k}1} ,\hat c_{\vec{k} 2} ,\cdots,\hat c_{\vec{k} N}\right)^T$
is the fermionic operator. $i=1,2,\cdots,N$ might denote
the spin of electrons, the sublattice index in the case of
a honeycomb lattice or the internal state of atoms.
The single-particle Hamiltonian $\mathcal{H}_{\vec{k}}$
has $N$ eigenvalues, which form $N$ energy bands, respectively.
If the Fermi energy lies within the band gap,
the bands lower than the Fermi energy are fully occupied
in the ground state, but those above the Fermi energy are empty.
The Chern number of the ground-state wave function is defined as
(see an introduction of the Chern number in Ref.~[\onlinecite{shen}])
\begin{equation}\label{eq:cherngeneral}
 C=\frac{i}{2\pi} \sum_{\alpha \in oc} \int d\vec{k}^2 \left( \Braket{ \frac{\partial 
 u_{\vec{k}\alpha}}{\partial k_x} | \frac{\partial u_{\vec{k}\alpha}}{\partial k_y}} - \text{H.c.}\right),
\end{equation}
where $|u_{\vec{k}\alpha}\rangle$ is the eigenvector of $\mathcal{H}_{\vec{k}}$
with $\alpha=1,2,\cdots,N$ denoting the different bands.
The sum of $\alpha$ is over all the occupied bands,
and the integral with respect to $\vec{k}$ is over the Brillouin zone.
The Chern number is a topological invariant, which is robust
against a local deformation of the Hamiltonian and can only take
an integer value.

The Chern number describes the topological property of the
ground-state wave function.
In the celebrated paper by Thouless {\it et} al.~\cite{TKNN},
the Chern number is related to the Hall conductance $\sigma_H^{G}$ as
\begin{equation}\label{eq:gsHall}
\sigma_H^{G}= Ce^2/h. 
\end{equation}
The Chern number for the ground state is then a measurable physical quantity.
Or one can say that the Hall conductance is the topological order parameter
of the ground state.
When the Chern number is nonzero,
the Fermi gas displays a quantum Hall effect even if there is no net
magnetic field~\cite{haldane}. The system is then called a Chern insulator.

Suppose that the system is initially in the ground state of
$\hat H_i$, before the Hamiltonian is quenched into $\hat H_f$.
The system is then driven out of equilibrium,
and the wave function follows a unitary time evolution
$|\Psi(t)\rangle = e^{-i \hat H_f t} |\Psi(0)\rangle$, where
$|\Psi(0)\rangle$ denotes the ground state of $\hat H_i$.
In a few paradigmatic models~\cite{alessiol,caio,wang15}, the Chern number of the unitarily-evolving
wave function is shown to be a constant, being independent of the post-quench
Hamiltonian $\hat H_f$. If one chooses it
as the topological order parameter of the quenched state, the topological order is
always robust against a quench of the Hamiltonian. However, the Chern number of
the unitarily-evolving wave function cannot be directly measured.
To address the topological order parameter
of the quenched state in an experimentally relevant way,
we study a measurable quantity - the Hall conductance.
In the quenched state, the Hall conductance must be distinguished
from the Chern number.

We notice that the Hall conductance cannot be
expressed as the expectation value of a local operator with respect to
the unitarily-evolving wave function $|\Psi(t)\rangle$.
Instead, it is the long-time response of the system
to an external electric field in linear response theory~\cite{mahan}.
This fact is related to the observation that measuring the Hall conductance
unavoidably introduces
decoherence and therefore in the long time limit
the system cannot be described by the unitarily-evolving wave function any more.
Instead, the system should be described by the diagonal ensemble~\cite{rigol08}.
If the expectation value of an observable $\hat O$ after the quench
relaxes to some steady value, it must be equal to its time average:
\begin{equation}
 \lim_{t\to\infty} O(t) = \lim_{T\to \infty} \frac{1}{T}\int^T_0 dt
\langle \Psi(t)| \hat O |\Psi(t)\rangle . 
\end{equation}
We insert the eigenbasis of $\hat H_f$ in the right-hand side.
In the limit $T\to\infty$, the off-diagonal terms of $\hat O$
are averaged out~\cite{rigol08}, we then have
\begin{equation}\label{eq:diaenarg}
\begin{split}
 \lim_{t\to\infty} O(t) & = \sum_{E}  | \langle E|\Psi(0)\rangle |^2 \langle E| \hat O |E\rangle \\
 & = \textbf{Tr} [\hat O \hat \rho],
\end{split}
 \end{equation}
where $|E\rangle$ is the eigenstate of $\hat H_f$ and $\hat \rho$ is diagonal in the basis $|E\rangle$
with the diagonal elements $| \langle E|\Psi(0)\rangle |^2$.
The diagonal ensemble $\hat \rho$ is obtained by dropping the off-diagonal terms
in the initial density matrix. This is equivalent to considering the decoherence effect.
Even though Eq.~(\ref{eq:diaenarg}) is based on the hypothesis of non-degenerate eigenenergies,
the diagonal ensemble is also applicable
in many integrable quantum systems~\cite{rigol09,ziraldo}.

Based on the above argument, the Hall conductance of the quenched state
should be calculated in the diagonal ensemble.
Let us represent an arbitrary many-body eigenstate of $\hat H_f$ by $|\{z_{\vec{k}\alpha} \}\rangle$,
where $z_{\vec{k}\alpha}=1,0$ denotes whether the single-particle state $|u^f_{\vec{k}\alpha}\rangle$
is occupied or not, respectively. Note that $|u^f_{\vec{k}\alpha}\rangle$
is the eigenvector of the post-quench single particle Hamiltonian $\mathcal{H}^f_{\vec{k}}$. The diagonal ensemble
can be written as
\begin{equation}\label{eq:diagonalensemblege}
 \hat \rho = \sum_{\{z_{\vec{k}\alpha} \}} p({\{z_{\vec{k}\alpha} \}}) |\{z_{\vec{k}\alpha} \}
 \rangle \langle \{z_{\vec{k}\alpha} \} |,
\end{equation}
where $p({\{z_{\vec{k}\alpha} \}})=|\braket{\{z_{\vec{k}\alpha} \}|\Psi(0)}|^2$
and the sum is over all the possible occupation configurations.

Now let us suppose a system located in the $x$-$y$ plane with its density
matrix being $\hat \rho$.
$\hat \rho$ is stationary in the sense that $[\hat \rho,\hat H_f]=0$.
Therefore, we can use the linear response theory
to calculate the Hall conductance by simply
replacing the thermal ensemble by $\hat \rho$.
In the linear response theory~\cite{mahan},
we suppose that an infinitesimal electric field is switched on in the $x$-direction,
and then the current in the $y$-direction is measured after infinitely long time.
The current is proportional to the electric field strength with
the constant of proportionality defined as
the Hall conductance.
The Hall conductance can be expressed by the current-current correlation as
\begin{equation}\label{eq:hallcond}
 \sigma_{H} = \lim_{\omega \to 0} \frac{1}{S\omega} \int^\infty_0 dt e^{i\omega t} \textbf{Tr}
 \left( \hat \rho  \left[ \hat J_y , \hat J_x(t) \right] \right),
\end{equation}
where $\omega$ denotes the frequency of the electric field and
the limit $\omega\to 0$ corresponds to the dc-conductance, and $S$
denotes the area of the system. $\hat J_x$
and $\hat J_y$ are the current operators in the $x$- and $y$-directions,
respectively. They are written as
\begin{equation}
\hat J_{x/y}= e \sum_{\vec{k}}
\hat c^\dag_{\vec{k}} \displaystyle \frac{\partial \mathcal{H}^f_{\vec{k}} }{\partial k_{x/y}}
\hat c_{\vec{k}}.
\end{equation}
Since both the Hamiltonian and the current operator are quadratic, we
can reexpress the Hall conductance by using the single-particle states as
\begin{equation}\label{eq:linearHall}
 \sigma_{H} = \lim_{\omega \to 0} \frac{1}{S\omega} \int^\infty_0 dt e^{i\omega t}
\sum_{\vec{k},\alpha} n_{\vec{k}\alpha} \langle u^f_{\vec{k}\alpha} | \left[ \hat J^y_{\vec{k}} , \hat J^x_{\vec{k}} (t) \right]
| u^f_{\vec{k}\alpha} \rangle,
\end{equation}
where the momentum-resolved current operator is a $N$-by-$N$ matrix, defined as
$\hat J^{x/y}_{\vec{k}}= e \displaystyle \frac{\partial \mathcal{H}^f_{\vec{k}} }{\partial k_{x/y}}$.
The occupation number $n_{\vec{k}\alpha}$
is related to the probability function $p({\{z_{\vec{k}\alpha} \}})$ by
$n_{\vec{k}\alpha}=\sum_{\{z_{\vec{k}'\alpha'} \}} p({\{z_{\vec{k}'\alpha'} \}}) \delta_{z_{\vec{k}\alpha},1}$.
In Eq.~(\ref{eq:linearHall}) the sum of $\alpha$ is over all the energy bands.

When calculating the integral with respect to $t$ in Eq.~(\ref{eq:linearHall}),
one can insert a factor $e^{-\eta|t|}$ into the integrand
with $\eta$ being an infinitesimal number. The integral then becomes convergent.
Because the dc Hall conductance must be a real number,
we keep the real part of $\sigma_H$, but neglect the imaginary part.
The Hall conductance becomes
\begin{equation}
\begin{split}
\sigma_H
 = & \frac{-i e^2}{S} \sum_{\vec{k},\alpha,\beta}\frac{n_{\vec{k}\alpha}}{(\epsilon_{\vec{k}\alpha}
 -\epsilon_{\vec{k}\beta})^2} \\ & \times \biggl[ \langle u^f_{\vec{k}\alpha} |
 \frac{\partial \mathcal{H}^f_{\vec{k}} }{\partial k_y} | u^f_{\vec{k}\beta}\rangle
 \langle u^f_{\vec{k}\beta} |
 \frac{\partial \mathcal{H}^f_{\vec{k}} }{\partial k_x} | u^f_{\vec{k}\alpha}\rangle - \text{H.c.} \biggr],
 \end{split}
\end{equation}
where $\epsilon_{\vec{k}\alpha}$ denotes the eigenvalue of $\mathcal{H}^f_{\vec{k}}$
in the band $\alpha$.
In thermodynamic limit, $\sum_{\vec{k}}$ is replaced by
$\frac{S}{(2\pi)^2} \int d\vec{k}^2 $.
By using the relation $\mathcal{H}^f_{\vec{k}} = \sum_\alpha \epsilon_{\vec{k}\alpha}
| u^f_{\vec{k}\alpha} \rangle\langle u^f_{\vec{k}\alpha} |$, we finally express the Hall conductance as
\begin{equation}
\begin{split}
\sigma_H = & \frac{e^2}{h} \frac{i}{2\pi} \sum_\alpha \int d\vec{k}^2 n_{\vec{k}\alpha} \left(\Braket{
\frac{\partial u^f_{\vec{k}\alpha} }{\partial k_x } | \frac{\partial u^f_{\vec{k}\alpha} }{\partial k_y } }
- \text{H.c.} \right).
\end{split}
\end{equation}
We choose $e^2/h$ as the unit of conductance.
The dimensionless Hall conductance $C_{neq}=\sigma_H/(e^2/h)$ is expressed as
\begin{equation}\label{eq:reducedHall}
\begin{split}
C_{neq} = & \frac{i}{2\pi} \sum_\alpha \int d\vec{k}^2 n_{\vec{k}\alpha} \left(\Braket{
\frac{\partial u^f_{\vec{k}\alpha} }{\partial k_x } | \frac{\partial u^f_{\vec{k}\alpha} }{\partial k_y } } 
- \text{H.c.} \right).
\end{split}
\end{equation}

This formula of Hall conductance stands for the
quenched states in general Fermi gases, in which the interaction between
fermions is neglected.
We notice that a Kubo formula calculation
by Dehghani, Oka and Mitra~\cite{dehghani} derived the similar formula
in a different way for Floquet topological states.
Comparing Eq.~(\ref{eq:reducedHall}) with Eq.~(\ref{eq:cherngeneral}),
we see the difference between the quench-state Hall conductance
and the ground-state Chern number. In the integrand of Eq.~(\ref{eq:reducedHall}),
the Berry curvature is weighted by the occupation number $n_{\vec{k}\alpha}$,
and the sum of $\alpha$ is over all the bands.
In the special case of $\hat H_i = \hat H_f$ (no quench),
the occupation is either $0$ for an empty band or $1$ for a
fully occupied band, and then the Hall conductance
reduces to the Chern number.
But for $\hat H_i\neq \hat H_f$,
no bands are fully occupied or completely empty.
$n_{\vec{k}\alpha}$ changes continuously with the momentum $\vec{k}$.
The integrand of Eq.~(\ref{eq:reducedHall})
cannot be expressed as the curl of some function
in the Brillouin zone, so that $C_{neq}$
is not quantized, but can take an arbitrary value.
This is different from the Chern number $C$, which must be an integer.

\section{Quench-state Hall conductance in a two-band Chern insulator}
\label{sec:halltwoband}

Eq.~(\ref{eq:reducedHall}) gives the Hall conductance of the quenched states
in $N$-component Fermi gases. Next we discuss
the case of $N=2$, in which the Hall conductance can be
conveniently expressed as a function of the parameters in $\hat H_i$
and $\hat H_f$ by utilizing the properties of the $SU(2)$ algebra.

In a two-component Fermi gas, the single-particle Hamiltonian can always be decomposed into
\begin{equation}\label{eq:hampauli}
\mathcal{H}_{\vec{k}}=\vec{d}_{\vec{k}} \cdot \vec{\sigma},
\end{equation}
where $\vec{\sigma}=(\sigma_x,\sigma_y,\sigma_z)$ denote the Pauli matrices.
There might be an additional constant in the expression of $\mathcal{H}_{\vec{k}}$;
however, the constant term has no effect on the eigenvectors, and, thus, does not contribute
to the Hall conductance. The Hall conductance is only determined by
the coefficient vectors $\vec{d}_{\vec{k}}=(d_{1\vec{k}},
d_{2\vec{k}},d_{3\vec{k}})$ in the initial and post-quench Hamiltonians.

The two eigenvalues of $\mathcal{H}_{\vec{k}}$
are $\pm d_{\vec{k}}$ with
\begin{equation}
 d_{\vec{k}} = \sqrt{ (d_{1\vec{k}})^2 + (d_{2\vec{k}})^2 + (d_{3\vec{k}} )^2} .
\end{equation}
The system has two bands, namely the lower band corresponding to the negative eigenvalue
and the upper band corresponding to the positive eigenvalue.
There is a gap between the two bands if we have $d_{\vec{k}}\neq 0$ everywhere in the Brillouin zone.

We set the chemical potential to zero,
in which case the lower band is fully occupied but the upper band is empty.
The Chern number of the ground state is written as
\begin{equation}\label{eq:cherntwobands}
 C=\frac{i}{2\pi} \int d\vec{k}^2 \left( \Braket{ \frac{\partial 
 u_{\vec{k}-}}{\partial k_x} | \frac{\partial u_{\vec{k}-}}{\partial k_y}} - \text{H.c.}\right),
\end{equation}
where $|u_{\vec{k}-}\rangle$ denotes the eigenvector of $\mathcal{H}_{\vec{k}}$
with the negative eigenvalue. The Chern number can be expressed by using
the coefficient vector as~\cite{shen}
\begin{equation}\label{eq:cherntwo}
 C= \int d\vec{k}^2 \frac{\left( \displaystyle\frac{\partial \vec{d}_{\vec{k}}}{\partial k_x}\times
 \frac{\partial \vec{d}_{\vec{k}}}{\partial k_y}\right) \cdot \vec{d}_{\vec{k}}}
 {4\pi d_{\vec{k}}^3}.
\end{equation}
The Chern number usually keeps invariant as the Hamiltonian changes.
But at some special points of the parameter space, the energy gap
closes ($d_{\vec{k}}=0$) somewhere in the Brillouin zone.
The Chern number then has a jump,
which indicates a topological phase transition in the ground state.

We quench the Hamiltonian
from $\mathcal{H}^i_{\vec{k}}= \vec{d}_{\vec{k}}^i\cdot \vec{\sigma}$
to $\mathcal{H}^f_{\vec{k}}= \vec{d}_{\vec{k}}^f\cdot \vec{\sigma}$,
and then calculate the Hall conductance in the quenched state.
The eigenvectors of $\mathcal{H}^{i/f}_{\vec{k}}$
with the positive and negative eigenvalues are
denoted by $\ket{u^{i/f}_{\vec{k}\alpha}}$ with $\alpha=\pm$, respectively.
The momentum $\vec{k}$
is a good quantum number in both the initial and the post-quench Hamiltonians.
Therefore, the occupation number is simply expressed as
\begin{equation}
 n_{\vec{k}\alpha} = |\langle u^f_{\vec{k}\alpha} | u^i_{\vec{k} -} \rangle |^2 ,
\end{equation}
where $| u^i_{\vec{k} -} \rangle$ is the initial state according to our protocol.
The total occupation at each $\vec{k}$ is conserved, satisfying
\begin{equation}\label{eq:occupationconserved}
n_{\vec{k}+}+ n_{\vec{k}-} \equiv 1.
\end{equation}
In fact, the occupation number evaluates
\begin{equation}\label{eq:occresult}
n_{\vec{k}\pm} = \frac{1}{2} \mp\frac{1}{2}
 \frac{\vec{d}^f_{\vec{k}} \cdot \vec{d}^i_{\vec{k}}}{{d}^f_{\vec{k}} {d}^i_{\vec{k}}},
\end{equation}
where ${d}^{i/f}_{\vec{k}}$ is the length of the coefficient vector $\vec{d}^{i/f}_{\vec{k}}$.
$\displaystyle \frac{\vec{d}^f_{\vec{k}} \cdot \vec{d}^i_{\vec{k}}}{{d}^f_{\vec{k}} {d}^i_{\vec{k}}}$
is called the occupation factor, which is just the cosine of the angle between the initial and the post-quench
coefficient vectors.

In the two-band Chern insulator, the Berry curvatures in the lower and upper bands
are opposite to each other everywhere in the Brillouin zone~\cite{shen}. We have
\begin{equation}
\frac{i}{2\pi} \left( \left\langle \frac{\partial 
 u^f_{\vec{k}\pm}}{\partial k_x} \right|\left. \frac{\partial u^f_{\vec{k}\pm }}{\partial k_y}\right\rangle - \text{H.c.}\right)
 = \mp \mathcal{C}_{\vec{k}},
\end{equation}
where
\begin{equation}\label{eq:twobandsberry}
\mathcal{C}_{\vec{k}}=\frac{\left( \displaystyle\frac{\partial \vec{d}^f_{\vec{k}}}{\partial k_x}\times
 \frac{\partial \vec{d}^f_{\vec{k}}}{\partial k_y}\right) \cdot \vec{d}^f_{\vec{k}}}
 {4\pi (d^f_{\vec{k}})^3}.
\end{equation}
Substituting Eq.~(\ref{eq:occresult}) and Eq.~(\ref{eq:twobandsberry}) into
Eq.~(\ref{eq:reducedHall}), we determine the Hall conductance as
\begin{equation}\label{eq:halltwoband}
  C_{neq} = \int d \vec{k}^2 
  \frac{\left(\vec{d}^f_{\vec{k}} \cdot \vec{d}^i_{\vec{k}} 
  \right) \left( \displaystyle\frac{\partial \vec{d}^f_{\vec{k}}}{\partial k_x}\times
 \frac{\partial \vec{d}^f_{\vec{k}}}{\partial k_y}\right) \cdot \vec{d}^f_{\vec{k}}}
 {4\pi {d}^i_{\vec{k}} (d^f_{\vec{k}})^4},
\end{equation}
where the integral is over the Brillouin zone.

\section{Nonanalytic behavior of the Hall conductance in
the Dirac model, the Haldane model and the Kitaev honeycomb model}
\label{sec:nonanalytic}

Suppose that there is a tunable parameter in the Hamiltonian~(\ref{eq:hampauli}),
namely $M$ without loss of generality (the physical meaning of
$M$ will be demonstrated in next section).
The vector $\vec{d}_{\vec{k}}$ is a function of $M$.
We use $M_i$ to denote the free parameter in the initial Hamiltonian,
and $M_f$ to denote that in the post-quench Hamiltonian.
Every pair $(M_i,M_f)$ determines a quenched state.
We further suppose that the initial state (or $M_i$) is fixed.
The quench-state Hall conductance $C_{neq}$ is then a function of $M_f$.
If the function $C_{neq}(M_f)$ is nonanalytic at some point,
namely at $M_f=M_f^c$,
we say that there is a nonequilibrium phase transition at $M_f=M_f^c$.
The term ``nonequilibrium phase transition'' comes from the fact
that this phase transition happens in the quenched state which
is far from equilibrium.

Let us discuss why $C_{neq}(M_f)$ can be
nonanalytic. In a generic model,
$\vec{d}^{i/f}_{\vec{k}}$ is an analytic
function of $\vec{k}$ and $M_{i/f}$. According to Eq.~(\ref{eq:halltwoband}),
both numerator and denominator of the integrand
are analytic functions. The denominator is the product
of ${d}^i_{\vec{k}}$ and $ (d^f_{\vec{k}})^4$. The former is nonzero everywhere
in the Brillouin zone, since
the initial state is usually chosen to be a gapped state.
But $M_f$ is a variable. We use $M_f^c$ to denote
the point at which the energy gap
in the post-quench Hamiltonian closes.
At $M_f\neq M_f^c$, $d^f_{\vec{k}}$
is nonzero at each $\vec{k}$. Therefore, the integrand in Eq.~(\ref{eq:halltwoband}) is
an analytic function of $\vec{k}$ and is bounded everywhere in the Brillouin zone.
$C_{neq}$ is then analytic at $M_f\neq M_f^c$.
However, at $M_f=M_f^c$, the energy gap of $\hat H_f$
closes somewhere in the Brillouin zone, namely at $\vec{k}=\vec{q}$
without loss of generality. We then have $d^f_{\vec{q}}=0$.
The integrand in Eq.~(\ref{eq:halltwoband}) is divergent at $\vec{k}=\vec{q}$.
We expect $C_{neq}(M_f)$ to be nonanalytic at $M_f=M_f^c$.
It is worth emphasizing that the nonanalyticity of $C_{neq}(M_f)$
can only be found at the gap closing points of the post-quench Hamiltonian,
at which the Chern number for the ground state of $\hat H_f$ changes.

In this section we will show that
$C_{neq}(M_f)$ in some models is in fact nonanalytic. We will discuss three models
with different symmetries and dispersion relations,
which are the Dirac model, the Haldane model, and the Kitaev model on a honeycomb lattice.

\subsection{The Dirac model}

In the Dirac model~\cite{shen}, the coefficient vector is expressed as
\begin{equation}\label{eq:dvecDirac}
 \vec{d}_{\vec{k}}=(k_x,k_y,M-B k^2),
\end{equation}
where $k^2=k_x^2+k_y^2$. The Dirac model has continuous translational
symmetry in real space.
The range of $\vec{k}$ is over the whole momentum plane,
which is treated as the Brillouin zone in our formalism.
It is straightforward to show by using Eq.~(\ref{eq:cherntwo}) that
the Chern number is
\begin{equation}\label{eq:diracchern}
 C=\displaystyle\frac{1}{2}\left( \textbf{sgn}(M)+ \textbf{sgn}(B)\right).
\end{equation}
A topological phase transition happens in the ground state
at $M=0$ or $B=0$. As $M=0$, the energy gap
closes at $\vec{k}=0$.

\begin{figure}[tbp]
\includegraphics[width=1.0\linewidth]{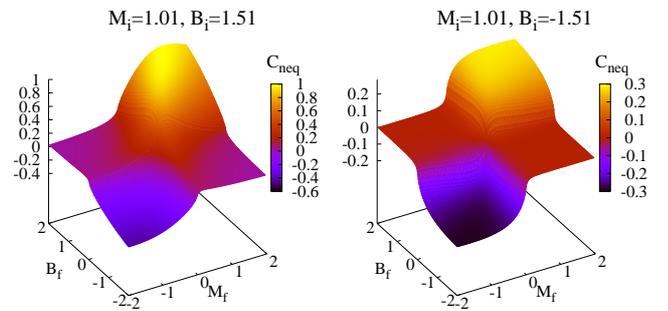}
\caption{(Color online) The Hall conductance $C_{neq}$ as a function of $(M_f,B_f)$ for
different $(M_i,B_i)$. [Left panel] The initial state
has a nonzero Chern number with $M_i,B_i>0$. [Right panel]
The Chern number of the initial state is zero with $M_i>0$ but $B_i<0$.}\label{fig:cneqfull}
\end{figure}
Substituting Eq.~(\ref{eq:dvecDirac}) into Eq.~(\ref{eq:halltwoband}), we obtain
the expression of the quench-state Hall conductance. It is
\begin{equation}\label{eq:halldiracint}
\begin{split}
 & C_{neq} \\ & = \int d\vec{k}^2 \frac{\left(
k^2 + (B_i k^2 -M_i)(B_f k^2 -M_f)\right)\left(B_f k^2 +M_f \right)}{4\pi {d}^i_{\vec{k}} (d^f_{\vec{k}})^4},
\end{split}
\end{equation}
with
\begin{equation}
{d}^i_{\vec{k}}=\sqrt{k^2+(B_i k^2 -M_i)^2},
\end{equation}
and
\begin{equation}\label{eq:df4dirac}
\left( {d}^f_{\vec{k}}\right)^4=\left(k^2+(B_f k^2 -M_f)^2 \right)^2.
\end{equation}
The Dirac model features rotational symmetry, which is
reflected in the fact that the integrand in Eq.~(\ref{eq:halldiracint})
is a function of $k^2$.
It is convenient to calculate this integral in the polar coordinates.
Integrating with respect to the azimuth angle, we obtain
\begin{equation}\label{eq:halldiracint2}
\begin{split}
 C_{neq} = & \int_0^\infty d\left({k}^2\right) \\ & \times \frac{\left(
k^2 + (B_i k^2 -M_i)(B_f k^2 -M_f)\right)\left(B_f k^2 +M_f \right)}{4 {d}^i_{\vec{k}} (d^f_{\vec{k}})^4}.
\end{split}
\end{equation}
The integrand in Eq.~(\ref{eq:halldiracint2})
is invariant as we simultaneously change the variable $k \to 1/k$ and
exchange the parameters $M_{i/f} \leftrightarrow B_{i/f}$.
The Hall conductance is then symmetric to $M_{i/f}$ and $B_{i/f}$ in the sense that
\begin{equation}\label{eq:diracsymm}
C_{neq}(M_i,B_i,M_f,B_f)= C_{neq}(B_i,M_i, B_f, M_f). 
\end{equation}

Fig.~\ref{fig:cneqfull} shows the result of the Hall conductance
obtained by numerical integration of Eq.~(\ref{eq:halldiracint2}).
$C_{neq}$ is plotted as a function
of the parameters $(M_f,B_f)$ in the post-quench Hamiltonian. The left
and right panels are for different initial states.
Different from the ground-state Hall conductance~(\ref{eq:diracchern}),
the quench-state Hall conductance changes continuously.
$C_{neq}$ is close to zero as $M_f$ and $B_f$ have different signs,
being positive as $M_f,B_f \gg 0$ but negative as $M_f,B_f \ll 0$.

The continuity and nonanalyticity of the function $C_{neq}(M_f,B_f)$
are proved as follows.
Since the denominator of the integrand in Eq.~(\ref{eq:halldiracint2})
contains an irrational function ${d}^i_{\vec{k}}$,
an analytical expression of $C_{neq}(M_f,B_f)$ is unaccessible.
But according to the argument at the beginning of this section, $C_{neq}(M_f,B_f)$
is nonanalytic at the gap closing point $M_f=0$.
As $M_f=0$, the gap of the post-quench Hamiltonian closes ($d^f_{\vec{k}}=0$)
at ${k}=0$, which is the unique singularity of the integrand in Eq.~(\ref{eq:halldiracint2}).
We then isolate this singularity by dividing the domain of integration into
\begin{equation}\label{eq:dividedomain}
 \int^\infty_0 d(k^2) =  \int^\eta_0 d(k^2) +  \int^\infty_\eta d(k^2),
\end{equation}
where $\eta>0$ can be arbitrarily small.
The integral $\int^\infty_\eta d(k^2)$ is an analytic function of $(M_f,B_f)$.
One can prove this by changing the variable $k\to 1/k$.
The domain of integration is changed into $[0,1/\eta]$, a finite interval.
At the same time, the new integrand is an analytic function
and is bounded everywhere in the domain $[0,1/\eta]$.
Therefore, the infinity is not a singularity of the integrand in Eq.~(\ref{eq:halldiracint2}).
We can neglect the second term in the right hand side of Eq.~(\ref{eq:dividedomain})
when analyzing the nonanalyticity of $C_{neq}$.

The nonanalytic part of $C_{neq}$ is now written as
\begin{equation}\label{eq:halldiracinteta}
\begin{split}
 C^\eta_{neq} = & \int_0^\eta d\left({k}^2\right) \\ & \times \frac{\left(
k^2 + (B_i k^2 -M_i)(B_f k^2 -M_f)\right)\left(B_f k^2 +M_f \right)}{4 {d}^i_{\vec{k}} (d^f_{\vec{k}})^4}.
\end{split}
\end{equation}
Since $\eta$ can be arbitrarily small, we replace the irrational function $ {d}^i_{\vec{k}}$
in the integrand by its value at $k=0$, i.e.,
\begin{equation}
 {d}^i_{\vec{k}=0} = |M_i|.
\end{equation}
This replacement does not change the nonanalytic behavior
of $C^\eta_{neq}$. Notice that we choose a gapped initial state and, thus, $M_i \neq 0$.
The integrand in the expression of $ C^\eta_{neq}$ becomes a rational function.
It is straightforward to determine $ C^\eta_{neq}$ as
\begin{equation}\label{eq:halldiracinteta}
\begin{split}
 C^\eta_{neq} = F(\eta)-F(0),
\end{split}
\end{equation} 
where $F$ denotes the antiderivative. $F(\eta)$
is an analytic function of $(M_f,B_f)$, while $F(0)$ is expressed as
\begin{widetext}
\begin{equation}\label{eq:fzeroexp}
 \begin{split}
  F(0)= \frac{1}{8|M_i| B^2_f }& \biggl[ \frac{2B_f(B_i+B_f)M_f -B_i }{\sqrt{1-4B_fM_f}} 
  \ln \frac{\left(2B_f^2M_f^2-4B_fM_f+1\right)\sqrt{1-4B_fM_f} - 8B_f^2M_f^2 +6B_fM_f -1}{M_f^2} \\ & 
  + B_i \ln M_f^2+2B_i-2B_f \biggr].
 \end{split}
\end{equation}
\end{widetext}
We are interested in the nonanalytic behavior of $C_{neq}^\eta$ at $M_f=0$.
In the vicinity of $M_f=0$, $\sqrt{1-4B_fM_f}$ can be expanded into
\begin{equation}
\begin{split}
 \sqrt{1-4B_fM_f} = & 1-2B_fM_f-2B_f^2M_f^2-4B_f^3M_f^3 \\ & 
 -10B_f^4M_f^4 + \mathcal{O} (M_f^5).
\end{split}
 \end{equation}
$F(0)$ is then expanded into
\begin{equation}\label{eq:fzeroexpexp}
 \begin{split}
  F(0)= & \frac{M_f}{4|M_i|}\ln M_f^2
  + \mathcal{O}(M_f^2) \ln | M_f| \\ &
  +\bigg[ \frac{2B_i-2B_f-B_i \ln(2B_f^4)}{8|M_i|B_f^2} +  \mathcal{O}(M_f) 
 \\ & - \frac{B_i}{8|M_i|B_f^2}
  \ln\left( 1+\mathcal{O}(M_f)\right) \\ &  + \mathcal{O}(M_f) \ln \left( 1+\mathcal{O}(M_f)\right) \bigg].
 \end{split}
\end{equation}

All the terms in Eq.~(\ref{eq:fzeroexpexp}) are continuous at $M_f= 0$.
Therefore, $F(0)$ and then $C_{neq}$ must be a continuous function even at $M_f = 0$.
The first and second terms in Eq.~(\ref{eq:fzeroexpexp}) are nevertheless
nonanalytic at $M_f=0$,
while all the other terms are analytic functions of $(M_f,B_f)$.
Since $F(\eta)$ and the difference between $C_{neq}$ and $C^\eta_{neq}$
are also analytic functions, we can then express the Hall conductance as
\begin{equation}\label{eq:noncneqdirac}
\begin{split}
 C_{neq} = & -\frac{1}{2} \frac{M_f-M_f^c}{|M_i-M_i^c|}\ln |M_f-M_f^c|
 \\ & + \mathcal{O}\left((M_f-M_f^c)^2\right) \ln | M_f-M_f^c|+ Ana.,
\end{split}
  \end{equation}
where $Ana.$ denotes an analytic function of $(M_f,B_f)$,
and $M_{i/f}^c=0$ denotes the gap closing point of the initial and post-quench Hamiltonian,
respectively.
The derivative of the first term in Eq.~(\ref{eq:noncneqdirac}) with respect to $M_f$
is divergent in the limit $M_f \to 0$.
But the second term in Eq.~(\ref{eq:noncneqdirac}) goes to zero faster than the first term,
and its derivative is finite at $M_f= 0$. We can then explicitly express the
nonanalytic behavior of $C_{neq}$ as
\begin{equation}
 \lim_{M_f\to M_f^c} \displaystyle\frac{\displaystyle\frac{\partial C_{neq}}{\partial M_f}}
 {-\displaystyle\frac{1}{2} \displaystyle\frac{\ln |M_f-M_f^c|}{|M_i-M_i^c|}} = 1,
\end{equation}
or simply write it as
\begin{equation}\label{eq:derivdirac}
 \lim_{M_f\to M_f^c} \frac{\partial C_{neq}}{\partial M_f} \sim
 {-\displaystyle\frac{1}{2} \displaystyle\frac{\ln |M_f-M_f^c|}{|M_i-M_i^c|}}.
\end{equation}

\begin{figure}[tbp]
\includegraphics[width=1.0\linewidth]{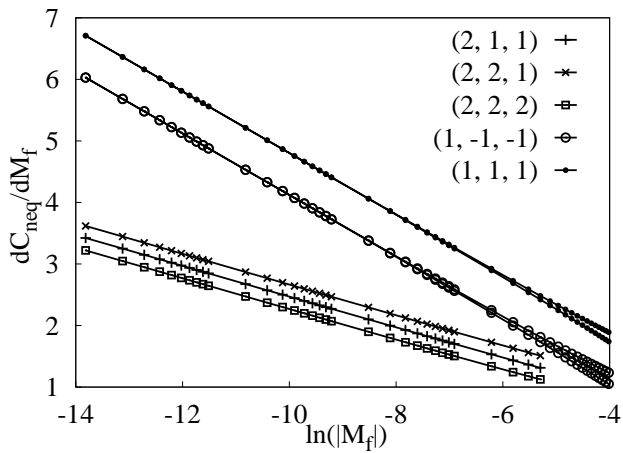}
\caption{${\partial C_{neq}}/{\partial M_f}$ as a function of $\ln |M_f|$.
The different types of lines and points represent ${\partial C_{neq}}/{\partial M_f}$
for different $(M_i,B_i,B_f)$.
We simultaneously plot ${\partial C_{neq}}/{\partial M_f}$
in the limit $M_f\to 0^+$ and $M_f \to 0^-$,
which are undistinguishable at small $|M_f|$.}\label{fig:diffcneq}
\end{figure}
In Fig.~\ref{fig:diffcneq}, we show the numerical result of $\frac{\partial C_{neq}}{\partial M_f}$
as a function of $\ln |M_f|$ in the vicinity of $M_f=0$.
In the range $|M_f|< e^{-6}$, this function is approximately linear
with the slope $-1/2|M_i|$. The numerical results verify
our analysis.

According to Eq.~(\ref{eq:diracsymm}), $C_{neq}$ is symmetric to $M_f$ and $B_f$.
By replacing $M_{i/f}$ by $B_{i/f}$ in Eq.~(\ref{eq:derivdirac}), we obtain
\begin{equation}
 \lim_{B_f\to B_f^c} \frac{\partial C_{neq}}{\partial B_f} \sim
 {-\displaystyle\frac{1}{2} \displaystyle\frac{\ln |B_f-B_f^c|}{|B_i-B_i^c|}},
\end{equation}
where $B_f^c=B_i^c=0$.
It is worth mentioning that the energy gap of the post-quench
Hamiltonian does not close at $B_f=0$.
The Hall conductance is nonanalytic at $B_f=0$ because
$k\to \infty$ becomes a singularity of the integrand in Eq.~(\ref{eq:halldiracint2}).
This is related to the fact that $k=0$ is a singularity at $M_f=0$,
since the integrand in Eq.~(\ref{eq:halldiracint2})
is invariant under the exchange $k\leftrightarrow 1/k$ and $M_{i/f}\leftrightarrow B_{i/f}$.
Whereas, in a generic model, the Brillouin zone is finite
so that a singularity at infinity does not exist. Therefore,
we can say that the Hall conductance can only be nonanalytic
at the gap closing point of the post-quench Hamiltonian.

The Hall conductance is continuous
everywhere in the parameter space of the quenched state.
But its derivative is logarithmically divergent at $M_f=0$ or $B_f=0$.
The nonanalyticity of the Hall conductance reveals a nonequilibrium phase transition
at $M_f=0$ or $B_f=0$.
Since the Chern number for the ground state
of the post-quench Hamiltonian is $\left(\textbf{sgn}(M_f)+\textbf{sgn} (B_f)\right)/2$,
this phase transition can be addressed by a change of topological invariant.
The nonanalytic behavior of the Hall conductance
at the nonequilibrium phase transition is quite
different from that at the ground-state phase transition. In the latter case,
the Hall conductance is discontinuous at the transition but its derivative keeps zero
almost everywhere in the parameter space.

This phase transition cannot be explained under the broken symmetry picture, since
the quenched states in different phases share the common symmetries of the Dirac model.
In fact, this phase transition is topologically driven.
And the Chern number for the ground state of the post-quench Hamiltonian
serves as a suitable order parameter, which can be used to distinguish
the quenched states in different phases (see Sec.~\ref{sec:topologicalinvariant}
for more discussion).

\subsection{The Haldane model}

\begin{figure}[tbp]
\includegraphics[width=1.0\linewidth]{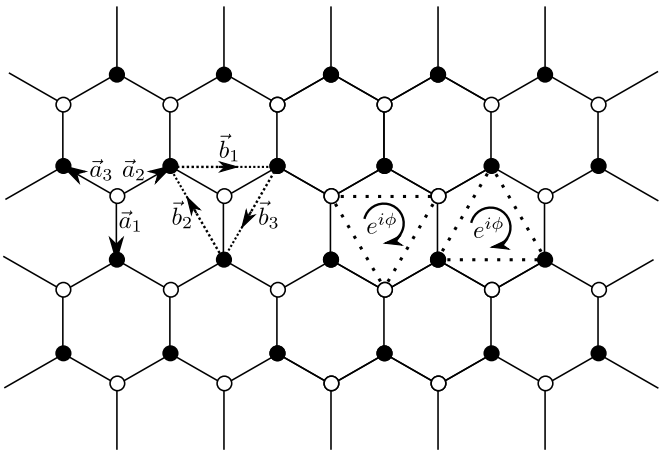}
\caption{Schematic diagram of the Haldane model. The black and
empty circles denote the ``$A$'' and ``$B$'' sites, respectively.
The circle arrow at the center of the dotted lines connecting
three ``$A$'' sites or three ``$B$'' sites shows the direction
of hopping with the matrix element $t_2e^{i\phi}$. The $6$ vectors
$\vec{a}_s$ and $\vec{b}_s$ with $s=1,2,3$ are marked, which are used for expressing
the Hamiltonian in momentum space.
}\label{fig:haldaneschematic}
\end{figure}
Next we study the quench-state Hall conductance in the Haldane model.
The model was first proposed by F.~D.~M.~Haldane~\cite{haldane} in 1988.
Due to the recent progress in manipulating cold atoms,
the Haldane model was realized in an optical lattice~\cite{jotzu}.
The study of the nonequilibrium phase transition
in the Haldane model provides an opportunity for testing our theory.

The Haldane model describes a Fermi gas on a honeycomb lattice
with each site at most being occupied by a single fermion.
Fig.~\ref{fig:haldaneschematic} is the schematic diagram.
There are two interpenetrating sublattices, which are the sublattice ``$A$'' denoted
by the black circles and the sublattice ``$B$''
denoted by the empty circles.
For simplicity, we set the lattice constant (the edge length of
the hexagon) to unity. In the Haldane model, the Hamiltonian
contains three terms:
\begin{equation}
 \hat H = \hat H_1 + \hat H_2 + \hat H_3.
\end{equation}
The first term describes the hopping between the nearest neighbors, i.e.,
between one ``$A$'' site and one ``$B$'' site, with the hopping matrix element set
to unity. $\hat H_1$ is expressed as
\begin{equation}
 \hat H_1 = \sum_{\langle \vec{A}_i,\vec{B}_j\rangle } \left(
 \hat c^\dag_{\vec{A}_i} \hat c_{\vec{B}_j} + \text{H.c.} \right),
\end{equation}
where $\hat c^\dag_{\vec{A}_i}$ and $\hat c_{\vec{B}_j}$ are the fermionic operators,
$\vec{A}_i$ and $\vec{B}_j$ denote
different ``$A$'' and ``$B$'' sites, respectively, and $\langle \vec{A}_i,\vec{B}_j\rangle$
denotes the nearest-neighbor relation. The second term describes the hopping
between the next-nearest neighbors, i.e., between two ``$A$'' sites or between
two ``$B$'' sites. The hopping matrix elements are complex numbers.
And inside each hexagon, it is $\left( t_2 e^{i\phi}\right)$ if the hopping is
in the clockwise direction (see the circle arrow in Fig.~\ref{fig:haldaneschematic}),
but $\left(t_2 e^{-i\phi}\right)$ if the hopping is in the anticlockwise direction.
$\hat H_2$ is expressed as
\begin{equation}
\begin{split}
 \hat H_2 = & \sum_{\langle \langle \vec{A}_i,\vec{A}_j\rangle\rangle } 
 \left( t_2e^{i\phi} \hat c^\dag_{\vec{A}_i} \hat c_{\vec{A}_j} + \text{H.c.} \right) \\
 & + \sum_{\langle \langle \vec{B}_i,\vec{B}_j\rangle\rangle } 
 \left( t_2e^{i\phi} \hat c^\dag_{\vec{B}_i} \hat c_{\vec{B}_j} + \text{H.c.} \right),
 \end{split}
\end{equation}
where $t_2$ and $\phi$ are real numbers, and $\langle \langle \vec{A}_i,\vec{A}_j\rangle\rangle$
denotes that the sites $\vec{A}_i$ and $\vec{A}_j$ are the next-nearest neighbors
to each other
and the hopping from $\vec{A}_j$ to $\vec{A}_i$ is in the clockwise direction.
The third term of the Hamiltonian describes an onsite potential which breaks the inversion symmetry.
$\hat H_3$ is expressed as
\begin{equation}
 \hat H_3 = M \sum_{\vec{A}_i}\hat c^\dag_{\vec{A}_i} \hat c_{\vec{A}_i}
 - M \sum_{\vec{B}_i}\hat c^\dag_{\vec{B}_i} \hat c_{\vec{B}_i}.
\end{equation}

The Haldane model is a two-band model. We express the Hamiltonian
in momentum space. In the basis
$\left( \hat c_{\vec{k}1}, \hat c_{\vec{k}2}\right)^T$ where
$\hat c_{\vec{k}1}= \sum_{\vec{A}_j} \frac{e^{-i\vec{k}\cdot \vec{A}_j}}{\sqrt{L}}
\hat c_{\vec{A}_j}$ and $\hat c_{\vec{k}2}= \sum_{\vec{B}_j} \frac{e^{-i\vec{k}\cdot \vec{B}_j}}{\sqrt{L}}
\hat c_{\vec{B}_j}$ ($L$ is the total number of sites), the single-particle Hamiltonian
is in the form of Eq.~(\ref{eq:hampauli}) with the components of $\vec{d}_{\vec{k}}$ expressed as
\begin{equation}\label{eq:vectordhaldane}
\begin{split}
d_{1\vec{k}} = &\sum_{s=1,2,3} \cos\left(\vec{k}\cdot \vec{a}_s\right), \\
d_{2\vec{k}} = & \sum_{s=1,2,3} \sin\left(\vec{k}\cdot \vec{a}_s\right), \\
d_{3\vec{k}} =& M-2t_2 \sin\phi \sum_{s=1,2,3} \sin \left(\vec{k}\cdot \vec{b}_s\right).
\end{split}
 \end{equation}
Here we employ $6$ constant vectors $\vec{a}_1, \vec{a}_2, \vec{a}_3$ and
$\vec{b}_1, \vec{b}_2, \vec{b}_3$, which are shown
in Fig.~\ref{fig:haldaneschematic}.

\begin{figure}[tbp]
\includegraphics[width=1.0\linewidth]{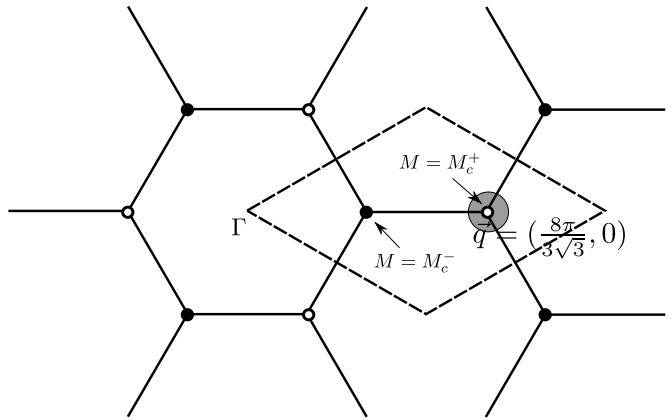}
\caption{The reciprocal lattice of the Haldane model. $\Gamma=(0,0)$
denotes the origin. The hexagon surrounding $\Gamma$ point is the
first Brillouin zone. The black (empty) circles denote
the singularities at which the energy gap closes for $M=M_c^-$ ($M=M_c^+$).
We choose the Brillouin zone surrounded by the dashed lines to calculate the Hall conductance.
$\vec{q}$ is the singularity inside this Brillouin zone as $M=M_c^+$,
and the shadow around $\vec{q}$ shows how we isolate this
singularity.}\label{fig:schematicbrillouin}
\end{figure}
In the Haldane model, the Chern number of the ground state is found to be
\begin{equation}
\begin{split}
 C=\frac{1}{2} \bigg( & \textbf{sgn}\left(M+3\sqrt{3} t_2\sin\phi \right) \\
& -\textbf{sgn}\left(M- 3\sqrt{3} t_2\sin\phi \right)\bigg).
 \end{split}
\end{equation}
Let us fix $t_2$ and $\phi$, while changing $M$. $M$ has two critical values, which are
\begin{equation}
M_c^\pm = \pm 3\sqrt{3} t_2\sin\phi.
\end{equation}
At $M=M_c^\pm$, the energy gap closes and the Chern number has a jump. One can obtain
the momentum $\vec{q}$ at which the gap closes by solving
the equation $d_{\vec{q}}=0$. Note that $\vec{q}$ is the singularity of the Berry curvature.
Since $\vec{d}_{\vec{k}}$ is a periodic function of $\vec{k}$ in the momentum plane,
the equation $d_{\vec{q}}=0$
has infinite number of solutions. As shown in Fig.~\ref{fig:schematicbrillouin},
for $M=M_c^+$ the energy gap closes at each empty circle,
while for $M=M_c^-$ the gap closes at each black circle.
The choice of the Brillouin zone is important for the calculation
of the Hall conductance which is an integral over the Brillouin zone.
We notice that, in the first Brillouin zone which is the hexagon centered at $\Gamma$
in Fig.~\ref{fig:schematicbrillouin}, the singularities
are located at the vertices. This causes some problem in the calculation.
We then choose a different Brillouin zone, which is the rhomboid
area surrounded by the dashed lines in Fig.~\ref{fig:schematicbrillouin}.
With this choice, the singularity is located inside the Brillouin zone.

Let us study the quench-state Hall conductance as a function of $M_f$
while fixing the initial parameters $M_i$, $t_{2i}$, and $\phi_i$ and the post-quench parameters
$t_{2f}$ and $\phi_f$. The Hall conductance is nonanalytic at $M_f=M_f^c$ with
\begin{equation}
M_f^c := + 3\sqrt{3} t_{2f}\sin\phi_f,
\end{equation}
where the gap of the post-quench Hamiltonian closes at the momentum
\begin{equation}
 \vec{q}= \left( \frac{8\pi}{3\sqrt{3}}, 0 \right).
\end{equation}
The Hall conductance is also nonanalytic at $M_f = - 3\sqrt{3} t_{2f}\sin\phi_f$.
But the nonanalytic behavior of the Hall conductance is the same at the two
different gap closing points. We will then only discuss the case of $M_f=M_f^c$.

Substituting Eq.~(\ref{eq:vectordhaldane}) into Eq.~(\ref{eq:halltwoband}),
we obtain
\begin{equation}\label{eq:inthallhaldane}
 C_{neq} = \int d\vec{k}^2 \frac{nu.}{4\pi d^i_{\vec{k}} \left( d^f_{\vec{k}} \right)^4 },
\end{equation}
where the numerator is expressed as
\begin{equation}
\begin{split}
 nu. = & \frac{\sqrt{3}}{2}\left( 3+2S_C + d^{i}_{3\vec{k}} d^{f}_{3\vec{k}}\right) \\
& \times \left( \frac{2}{3\sqrt{3}} M_f^c \sum_{s} \left(\cos(\vec{k}\cdot \vec{b}_s) -1 \right)^2 -M_f S_S \right).
\end{split}
\end{equation}
In the denominator, we have
\begin{equation}
 d^i_{\vec{k}} = \sqrt{\left(M_i-\frac{2}{3\sqrt{3}}M_i^c S_S \right)^2+3+2S_C},
\end{equation}
with $M_i^c=3\sqrt{3} t_{2i} \sin \phi_i$ denoting the gap closing point
of the initial Hamiltonian and
\begin{equation}\label{eq:df4haldane}
 \left( d^f_{\vec{k}} \right)^4 = \left(\left(M_f-\frac{2}{3\sqrt{3}}M_f^c S_S \right)^2+3+2S_C\right)^2.
\end{equation}
Here the symbols $S_C$ and $S_S$ denote the summations
\begin{equation}
 S_C = \sum_s \cos \left( \vec{k}\cdot \vec{b}_s \right),
\end{equation}
and
\begin{equation}
 S_S = \sum_s \sin \left( \vec{k}\cdot \vec{b}_s \right),
\end{equation}
respectively. $d^{i}_{3\vec{k}}$ and $d^{f}_{3\vec{k}}$ are the third components of
the initial coefficient vector $\vec{d}^{i}_{\vec{k}}$ and the post-quench
coefficient vector $\vec{d}^{f}_{\vec{k}}$, respectively.

An analytical expression of $C_{neq}(M_f)$ is unaccessible.
We analyze the nonanalyticity of $C_{neq}(M_f)$ by using the same trick
as we used for the Dirac model.
We divide the Brillouin zone into a circle of infinitesimal radius
centered at the singularity $\vec{q}$ (the shadow in Fig.~\ref{fig:schematicbrillouin})
and the left area. The nonanalyticity of $C_{neq}(M_f)$ comes only from
the integral in the vicinity of the singularity, which is written as
\begin{equation}\label{eq:cetahaldane}
C^\eta_{neq}= \int_{\mathcal{B}_\eta(\vec{q})} d\vec{k}^2 \frac{nu.}{4\pi d^i_{\vec{k}} \left( d^f_{\vec{k}} \right)^4 },
\end{equation}
where $\mathcal{B}_\eta(\vec{q})$ denotes a circle of radius $\sqrt{\eta}$
centered at $\vec{q}$ with $\eta$ being an infinitesimal number.
While $\left(C_{neq}-C^\eta_{neq}\right)$ is an analytic function of $M_f$,
since the integrand in Eq.~(\ref{eq:inthallhaldane}) is an analytic function of $M_f$ and $\vec{k}$
and is bounded for $\vec{k}\notin \mathcal{B}_\eta(\vec{q})$.
The domain of integration for $C^\eta_{neq}$ can be arbitrarily small.
Therefore, in the denominator of the integrand in Eq.~(\ref{eq:cetahaldane})
we replace $d^i_{\vec{k}}$ by its value at $\vec{q}$, i.e.,
\begin{equation}\label{eq:diexphaldane}
 d^i_{\vec{q}}=|M_i-M_i^c|.
\end{equation}
There are trigonometric functions in $\left( d^f_{\vec{k}} \right)^4$ which
prevent us from working out $C^\eta_{neq}$. Notice that we cannot replace $\left( d^f_{\vec{k}} \right)^4$
by its value at $\vec{k}=\vec{q}$ which vanishes at the gap closing point $M_f=M_f^c$.
But we can do an expansion of $\left( d^f_{\vec{k}} \right)^4$
in the vicinity of $\vec{k}=\vec{q}$. We set $\Delta \vec{k} = \vec{k}-\vec{q}$
and find
\begin{equation}\label{eq:Sexpansionhaldane}
\begin{split}
 S_C= & -\frac{3}{2} + \frac{9}{8} \Delta k^2 + \mathcal{O} (\Delta k^3),\\
 S_S = & \frac{3\sqrt{3}}{2} - \frac{9\sqrt{3}}{8} \Delta k^2 + \mathcal{O} (\Delta k^3),
 \end{split}
\end{equation}
where $\Delta k^2=\Delta k_x^2+\Delta k_y^2$ and $\mathcal{O} (\Delta k^3)$
denotes the higher-order terms. Substituting Eq.~(\ref{eq:Sexpansionhaldane})
into Eq.~(\ref{eq:df4haldane}), we obtain
\begin{equation}\label{eq:df4haldaneexp}
\begin{split}
 \left( d^f_{\vec{k}} \right)^4 =  \bigg[\frac{9}{4}\Delta k^2 
 + \left( \frac{3}{4}M_f^c \Delta k^2 + M_f-M_f^c 
 \right)^2  \bigg]^2.
 \end{split}
\end{equation}
For the Haldane model $ \left( d^f_{\vec{k}} \right)^4$
has the same form in the vicinity of the singularity as
for the Dirac model (see Eq.~(\ref{eq:df4dirac})).
Notice that in the Dirac model we have $\Delta k^2=k^2$ due to
$\vec{q}=0$ (the singularity is at the origin), and $M_f-M_f^c=M_f$ due to $M_f^c=0$.
This similarity indicates that the asymptotic behavior of the spectrum $d^f_{\vec{k}}$
nearby the singularity is universal in two-band Chern insulators.
We then expect that the nonanalytic behavior of the Hall conductance
in the Haldane model is similar to that in the Dirac model.

We can keep the expansion of $\left( d^f_{\vec{k}} \right)^2$
only to the second order and obtain
 \begin{equation}\label{eq:df2exphaldane}
\begin{split}
 \left( d^f_{\vec{k}} \right)^2 = \left(M_f-M_f^c\right)^2 +
\gamma\Delta k^2,
 \end{split}
\end{equation}
where
\begin{equation}
 \gamma=  \frac{3\left( M_f-M_f^c\right)M_f^c }{2}+\frac{9}{4} .
\end{equation}
The nonanalytic behavior of $C^\eta_{neq}(M_f)$
is independent of whether we use Eq.~(\ref{eq:df2exphaldane}) or Eq.~(\ref{eq:df4haldaneexp})
in the calculation.
A linear term is absent in Eq.~(\ref{eq:df2exphaldane}).
This reflects the conic structure of the spectrum at the singularities.

We expand the numerator to the second order in $\Delta k$ and obtain
\begin{equation}\label{eq:nuexphaldane}
\begin{split}
 nu. = \kappa_1 + \kappa_2 \Delta k^2,
\end{split}
 \end{equation}
where $\kappa_1$ and $\kappa_2$ are $\Delta k$-independent
constants with
\begin{equation}
\begin{split}
 \kappa_2 = & \frac{27}{8} \left( M_f-M_f^c\right) \\ & \times \left( \frac{1}{2}
 \left(M_f-M_f^c\right)M_i -\left(M_f-M_f^c\right)M_i^c-\frac{3}{2} \right).
 \end{split}
\end{equation}
In fact, only the second-order term $\kappa_2 \Delta k^2$
in the numerator contributes to the nonanalytic behavior of $C^\eta_{neq}$
at $M_f= M_f^c$, while $\kappa_1$ and all the higher-order
terms have no contribution and can be neglected.

Substituting Eq.~(\ref{eq:diexphaldane}), (\ref{eq:df2exphaldane}) and (\ref{eq:nuexphaldane})
into Eq.~(\ref{eq:cetahaldane}), we express the integrand for $C^\eta_{neq}$ as a function of
$\Delta k^2$. Integrating with respect to the azimuth angle in the polar coordinates,
we obtain
\begin{equation}
\begin{split}
& C^\eta_{neq} \\ & = \int_0^\eta d\left(\Delta{k}^2\right) \frac{\kappa_1+ \kappa_2 \Delta k^2}{4 |M_i-M_i^c|
\left( \left(M_f-M_f^c\right)^2 +
\gamma\Delta k^2 \right)^2 }.
\end{split}
\end{equation}
The calculation of this integral is straightforward.
Notice that we are only interested in the nonanalytic part of $C^\eta_{neq}$, which is
\begin{equation}\label{eq:noncneqhaldane}
 C^\eta_{neq} \sim \frac{-\kappa_2}{4|M_i-M_i^c|\gamma^2} \ln \left(M_f-M_f^c\right)^2.
\end{equation}
Furthermore, at $M_f=M_f^c$, $\kappa_2/\gamma^2$ can be expanded into
\begin{equation}
 \frac{\kappa_2}{\gamma^2} = -\left(M_f-M_f^c\right) + \mathcal{O}\left(M_f-M_f^c\right)^2.
\end{equation}
Finally, we express the Hall conductance as
\begin{equation}
\begin{split}
 C_{neq} = & \frac{\left(M_f-M_f^c\right)}{2|M_i-M_i^c|} \ln |M_f-M_f^c| \\ & + 
\mathcal{O}\left( \left(M_f-M_f^c\right)^2 \right) \ln |M_f-M_f^c| + Ana.,
\end{split}
\end{equation}
where $Ana.$ represents an analytic function of $M_f$.

\begin{figure}[tbp]
\includegraphics[width=1.0\linewidth]{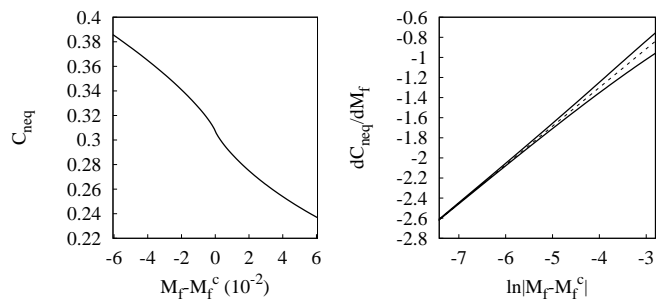}
\caption{The Hall conductance and its derivative in the vicinity of
$M_f=M_f^c$. The former is plotted in the left panel,
while the latter is plotted in the right panel. In the right panel, the solid lines
denote $\frac{d C_{neq}}{d M_f}$ obtained from numerical integration, while the dashed one is a straight line
with slope $\frac{1}{2|M_i-M_i^c|}$. The parameters are set to $M_i=0$, $\phi_i=\phi_f=\pi/6$,
$t_{2i}=0.5$, and $t_{2f}=1$.}\label{fig:hallhaldane}
\end{figure}
First, the Hall conductance $C_{neq}(M_f)$ is continuous everywhere, even at
$M_f=M_f^c$. Second, in the limit $M_f\to M_f^c$,
the derivative of the Hall conductance displays
the following asymptotic behavior:
\begin{equation}\label{eq:derivhaldane}
\displaystyle\frac{d C_{neq}}{d M_f} \sim 
\frac{1}{2} \displaystyle\frac{\ln |M_f-M_f^c|}{|M_i-M_i^c|} .
\end{equation}
We obtain the exact asymptotic behavior of $dC_{neq}/dM_f$.
We truncated the numerator and denominator
of the integrand when calculating the Hall conductance~(\ref{eq:inthallhaldane}),
but the higher-order terms that we neglected have no contribution to the first-order derivative
of $C_{neq}$ in the limit $M_f\to M_f^c$.
The numerical results verify Eq.~(\ref{eq:derivhaldane}). Fig.~\ref{fig:hallhaldane}
shows the Hall conductance and its derivative.
We see clearly that $C_{neq}$ is a continuous function and $d C_{neq}/d M_f$
is a linear function of $\ln |M_f-M_f^c|$ in the limit $M_f\to M_f^c$.
The slope of $d C_{neq}/d M_f$ coincides with our prediction.

Even if the Haldane model is significantly distinguished
from the Dirac model in symmetries and dispersion relations,
the derivative of the Hall conductance has the similar asymptotic behavior
except for a sign difference in these two models.
In both models, the Hall conductance is continuous everywhere
with a logarithmically-divergent derivative at the gap closing point
of the post-quench Hamiltonian.
And the prefactor of the logarithmically-divergent derivative
is inversely proportional to the energy gap in the initial state, i.e., $2|M_i-M_i^c|$.
The nonanalyticity of the Hall conductance is always accompanied by
the change of the Chern number for the ground state of the post-quench Hamiltonian.

\subsection{The Kitaev honeycomb model}

\begin{figure}[tbp]
\includegraphics[width=1.0\linewidth]{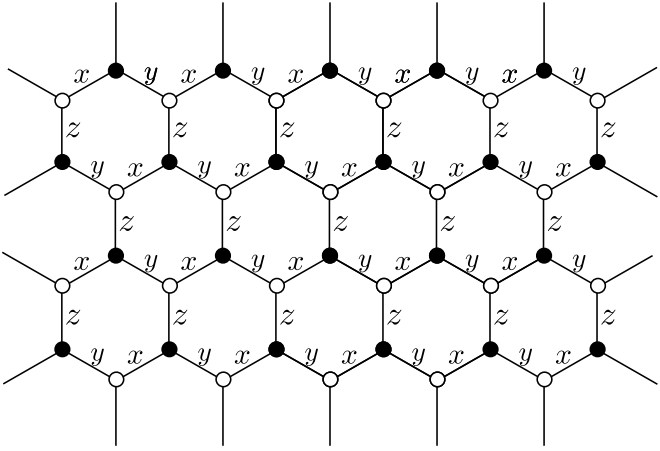}
\caption{Lattice of the Kitaev honeycomb model. Spin-1/2 degrees of freedom reside
on the vertices of a honeycomb lattice. The anisotropic nearest-neighbor interaction
depends on the link type ($x$, $y$, or $z$).}\label{fig:kitaev}
\end{figure}
Finally, we study the Kitaev honeycomb model in a magnetic field.
Different from the Dirac model or the Haldane model, the Kitaev honeycomb model~\cite{kitaev06}
is not a fermionic model, but a spin-$1/2$ model defined on a honeycomb lattice
with anisotropic nearest-neighbor interactions (see Fig.~\ref{fig:kitaev}).
The Hamiltonian of the Kitaev honeycomb model is
\begin{equation}
 \hat H_1 = -J_x\sum_{x-\text{links}} \sigma^x_i \sigma^x_j  
 - J_y \sum_{y-\text{links}} \sigma^y_i \sigma^y_j  
 - J_z \sum_{z-\text{links}} \sigma^z_i \sigma^z_j,  
\end{equation}
where the sum over the $x$, $y$ or $z$-links means the sum
over pairs of lattice sites $\langle i,j\rangle$ that are linked
by a bond labeled by $x$, $y$, or $z$ in Fig.~\ref{fig:kitaev}, respectively.
We set the bond length to unity.
It has been shown~\cite{chen08} that for the above Hamiltonian one can find a
Jordan-Wigner contour, which after identifying a conserved $Z_2$ operator~\cite{kitaev06}
and switching to momentum space yields a BCS-type Hamiltonian:
\begin{equation}
 \hat H_1 = \sum_{\vec{k}} \left( \frac{ \varepsilon_{\vec{k}}}{2}\left( \hat c^\dag_{\vec{k}}
 \hat c_{\vec{k}} - \hat c_{-\vec{k}} \hat c^\dag_{-\vec{k}}\right)
 + \frac{\Delta_{\vec{k}}}{2} \left(\hat c^\dag_{\vec{k}} \hat c^\dag_{-\vec{k}}
 + \hat c_{-\vec{k}} \hat c_{\vec{k}}\right) \right)
\end{equation}
with
\begin{equation}
\begin{split}
 \varepsilon_{\vec{k}}=2\bigg[ & J_z + J_x \cos\left( \frac{\sqrt{3}k_x}{2}
 + \frac{3k_y}{2} \right) \\ & + 
 J_y \cos\left( -\frac{\sqrt{3}k_x}{2}
 + \frac{3k_y}{2} \right)\bigg], \\
 \Delta_{\vec{k}} = 2 \bigg[& J_x \sin\left( \frac{\sqrt{3}k_x}{2}
 + \frac{3k_y}{2} \right) \\ & + J_y \sin\left( -\frac{\sqrt{3}k_x}{2}
 + \frac{3k_y}{2} \right)\bigg].
 \end{split}
\end{equation}
This Hamiltonian can be diagonalized by a Bogoliubov transformation
with a spectrum $\sqrt{\varepsilon_{\vec{k}}^2+\Delta_{\vec{k}}^2}$.
Analyzing this spectrum one finds a gapless phase
for $|J_{x}|<|J_{y}|+|J_{z}|$, $|J_{y}|<|J_{x}|+|J_{z}|$,
and $|J_{z}|<|J_{x}|+|J_{y}|$.

The presence of a magnetic field $\vec{h}$ adds an additional term to
the Hamiltonian, which is
\begin{equation}
 \hat H_2 = \sum_j \left( h_x \hat \sigma^x_j + h_y \hat \sigma^y_j
 + h_z \hat \sigma^z_j\right).
\end{equation}
The external field opens a gap in the gapless phase.
At $J_x=J_y=J_z=J$ there exists a diagonal form of the Hamiltonian also with
nonzero magnetic field~\cite{kitaev06} and the spectrum becomes $\sqrt{\varepsilon_k^2
+ |\tilde \Delta_{\vec{k}}|^2}$ with
\begin{equation}
\begin{split}
 \tilde \Delta_{\vec{k}} = & \Delta_{\vec{k}} + i 4M \bigg[
 \sin\left( \frac{\sqrt{3}k_x}{2}+\frac{3k_y}{2} \right)\\ &
 - \sin\left( -\frac{\sqrt{3}k_x}{2}+\frac{3k_y}{2} \right)
 - \sin\left( \sqrt{3}k_x \right)\bigg],
 \end{split}
\end{equation}
where $M\sim \frac{h_xh_yh_x}{J^2}$. The diagonal Hamiltonian can be transformed to
the two-band form in Eq.~(\ref{eq:hampauli}) via a Bogoliubov
transformation~\cite{schmitt}, yielding
\begin{equation}\label{eq:dveckitaev}
 \vec{d}_{\vec{k}}=\frac{1}{2} \left(\textbf{Re}\tilde \Delta_{\vec{k}},
 \textbf{Im}\tilde \Delta_{\vec{k}}, \varepsilon_{\vec{k}} \right).
\end{equation}
Without loss of generality we set $J=1$ in the following.

\begin{figure}[tbp]
\includegraphics[width=1.0\linewidth]{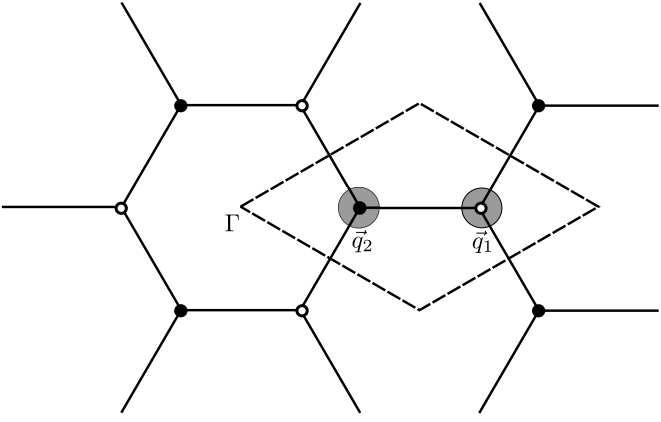}
\caption{The reciprocal lattice of the Kitaev honeycomb model. $\Gamma=(0,0)$
denotes the origin. The black and the empty circles denote
the singularities at which the energy gap closes as there is no magnetic field.
The dashed lines surround the Brillouin zone that we choose to calculate
the Chern number and the Hall conductance.
$\vec{q_1}$ and $\vec{q}_2$ are two singularities inside this Brillouin zone.
The shadows around them show how we separate the
singularities.}\label{fig:kitaevbrillouin}
\end{figure}
The spin-$1/2$ model with the Hamiltonian $\hat H=\hat H_1+\hat H_2$
is now transformed into a two-band model of
fermions. We can then define the Chern number.
The Chern number of the ground state is
\begin{equation}
C=\textbf{sgn} \left( M \right).
\end{equation}
A topological phase transition in the ground state happens at $M=0$, i.e., at zero magnetic field,
at which the energy gap closes with the roots of $d_{\vec{k}}$
sitting on the corners of the hexagonal Brillouin zone (see Fig.~\ref{fig:kitaevbrillouin}).
These roots are the singularities of the Berry curvature.
There are two types of conic singularities, denoted by
the black and the empty circles in Fig.~\ref{fig:kitaevbrillouin}.
As in the Haldane model, we choose a rhomboid unit cell
as the Brillouin zone.
But different from the Haldane model, the Brillouin zone in
the Kitaev model contains two singularities,
namely
\begin{eqnarray}
\begin{array}{cc}
 \vec{q}_1 = \left( \displaystyle\frac{8\pi}{3\sqrt{3}}, 0\right), & 
  \vec{q}_2 = \left( \displaystyle\frac{4\pi}{3\sqrt{3}}, 0 \right).
 \end{array}
\end{eqnarray}

Next we study the quench-state Hall conductance in
the two-component Fermi gas with the coefficient vector given by Eq.~(\ref{eq:dveckitaev}).
It is worth mentioning that the value of the $Z_2$ operator defined
in the transformation from the Kitaev model to the fermionic model is
conserved in a quench of the parameter $M$. Therefore,
a quench in the Kitaev model can be mapped
into a quench in the corresponding two-component Fermi gas and vice versa.
Whereas, the observable in the Kitaev model that corresponds
to the Hall conductance is difficult to write down, which will not
be discussed in this paper.

The Hall conductance depends on
the parameters $M_i$ and $M_f$ in the initial and post-quench Hamiltonians, respectively.
We fix $M_i$ and study the nonanalytic behavior of the function
$C_{neq}(M_f)$ in the limit $M_f \to 0$. According to Eq.~(\ref{eq:halltwoband}),
the nonanalyticity
of $C_{neq}(M_f)$ comes from the integral over the neighborhood of the singularities
$\vec{q}_1$ and $\vec{q}_2$ (the shadowed circles in Fig.~\ref{fig:kitaevbrillouin}).
An analysis of the symmetry properties of the integrand yields that
the integrand as a whole is invariant under $\vec{k}\to -\vec{k}$.
Therefore, the contributions to the integral in Eq.~(\ref{eq:halltwoband}) from both singularities
are identical and it suffices to consider
only one of them and then double the result.

We choose to consider a surrounding of $\vec{q}_1$ with its contribution
to the Hall conductance written as
\begin{equation}\label{eq:hallexphaldane}
C^\eta_{neq}=  
\int_{\mathcal{B}_\eta(\vec{q}_1)} d\vec{k}^2
\frac{\left(\vec{d}^f_{\vec{k}} \cdot \vec{d}^i_{\vec{k}} 
  \right) \left( \displaystyle\frac{\partial \vec{d}^f_{\vec{k}}}{\partial k_x}\times
 \frac{\partial \vec{d}^f_{\vec{k}}}{\partial k_y}\right)
 \cdot \vec{d}^f_{\vec{k}}}{4\pi d^i_{\vec{k}} \left( d^f_{\vec{k}} \right)^4 },
\end{equation}
where $\mathcal{B}_\eta(\vec{q}_1)$ denotes a circle centered at $\vec{q}_1$
with the radius $\sqrt{\eta}$. $\eta$ can be arbitrarily small.
In the denominator of the integrand in Eq.~(\ref{eq:hallexphaldane}),
$d^i_{\vec{k}}$ can be replaced by its value at $\vec{k}=\vec{q}_1$,
i.e., $d^i_{\vec{q}_1}=3\sqrt{3} |M_i|$.
We then expand the numerator and denominator of the integrand around $\vec{k}=\vec{q}_1$.
With $\Delta \vec{k}=\vec{k}-\vec{q}_1$ we obtain for the numerator
\begin{equation}\label{eq:expnukitaev}
\begin{split}
 nu. =  \kappa_1 + \kappa_2 \Delta k^2 
 +\kappa'_1 \Delta k_x + \kappa'_2 \Delta k_x^2  + \mathcal{O}(\Delta k^3)
\end{split}
 \end{equation}
with $\kappa_1=\frac{729\sqrt{3}}{4} M_i M_f^2$, $\kappa'_1=-\frac{729M_iM_f^2}{4}$ and
\begin{equation}
\begin{split}
 \kappa_2 = & M_f\left( \frac{243\sqrt{3}}{16}
 -\frac{2187\sqrt{3}}{16} M_i M_f \right), \\ 
 \kappa'_2 = & -\frac{2187\sqrt{3}}{8}M_iM_f^2.
 \end{split}
\end{equation}
For the denominator we obtain
\begin{equation}
 \left( d^f_{\vec{k}} \right)^2= 27 \left( M_f^2+ \gamma \Delta k^2 \right),
\end{equation}
where $\gamma=\frac{1}{12}- \frac{3M_f^2}{2}$.
The expansion of $\left( d^f_{\vec{k}} \right)^2$ reveals
the rotational symmetry in the conic structure of the spectrum nearby the singularities.
The similar structure has already been found in the Haldane model and in the Dirac model.

We consider the terms in the expansion of the numerator~(\ref{eq:expnukitaev}) one by one.
The constant $\kappa_1$
contributes to $C^\eta_{neq}(M_f)$ a regular term in the sense that
it is a continuous function of $M_f$ with the derivative
being finite at $M_f=0$.
The second-order term $\kappa_2 \Delta k^2$ in the numerator
contributes to $C^\eta_{neq}(M_f)$ a term $\sim M_f \ln \left|M_f\right|$,
the derivative of which is logarithmically divergent in the limit $M_f\to 0$.
There are two additional terms ($\kappa'_1\Delta k_x$ and $\kappa'_2\Delta k_x^2$)
in the numerator which break the rotational symmetry.
The contribution to $C^\eta_{neq}$ from these asymmetric terms is regular.
Terms linear in $\Delta k_\alpha$ (or to any odd power of $\Delta k_\alpha$)
are anti-symmetric under $\Delta k_\alpha \to -\Delta k_\alpha$.
While the denominator of the integrand are symmetric under $\Delta k_\alpha \to -\Delta k_\alpha$.
The integral of an anti-symmetric function must vanish. Therefore,
$\kappa'_1\Delta k_x$ has no contribution to $C^\eta_{neq}$.
On the other hand, since $\kappa'_2$ is proportional to $M_f^2$,
$\kappa'_2 \Delta k_x^2$ in the numerator of the integrand
contributes to $C^\eta_{neq}$ a nonanalytic
term $\sim M_f^2 \ln \left|M_f\right|$ which is regular.

Due to the above analysis the nonanalytic part of $C^\eta_{neq}$ is expressed as
\begin{equation}
\begin{split}
 C^\eta_{neq}\sim & \int^\eta_0 d \left( \Delta k^2\right)
 \frac{\kappa_2 \Delta k^2}
 {4 d^i_{\vec{q}_1} \left[ 27 \left( M_f^2+ \gamma \Delta k^2 \right)\right]^2 }
 \\ \sim & \frac{-M_f}{2|M_i|} \ln |M_f|  +
 \mathcal{O}\left( M_f^2 \right)\ln |M_f| .
 \end{split}
\end{equation}
Since there are two singularities
in the Brillouin zone, the Hall conductance can be expressed as
\begin{equation}
\begin{split}
 C_{neq} = & \frac{-(M_f-M_f^c)}{|M_i-M_i^c|} \ln |M_f-M_f^c| \\ & +
 \mathcal{O}\left( (M_f-M_f^c)^2 \right)\ln |M_f-M_f^c| + Ana.,
 \end{split}
\end{equation}
where $M_i^c=0$ and $M_f^c=0$ denote the gap closing point in the
initial and post-quench Hamiltonians, respectively, and $Ana.$ represents
an analytic function of $M_f$.
The Hall conductance is a continuous function of $M_f$, but its derivative
is divergent in a logarithmic way. And the asymptotic behavior
of the derivative at the gap closing point $M^c_f=0$ can be expressed as
\begin{equation}
 \lim_{M_f \to M_f^c} \frac{d C_{neq}}{d M_f} \sim \frac{-\ln|M_f-M_f^c|}{|M_i-M_i^c|}.
\end{equation}

\begin{figure}[tbp]
\includegraphics[width=1.0\linewidth]{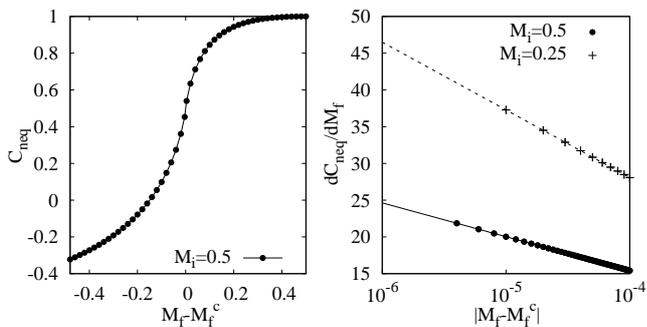}
\caption{[Left panel] The function $C_{neq}(M_f)$ for a quench starting from $M_i=1/2$
obtained by numerical integration (adaptive Simpson's) of Eq.~(\ref{eq:halltwoband})
over a half unit cell and doubling the result. [Right panel] Derivative of
$C_{neq}$ with respect to $M_f$ for a quench from $M_i=1/2$ and one quench from
$M_i=1/4$. The dots denote the numerical results, while the lines are fits of the form
$\left(-|M_i-M_i^c|^{-1}\ln |M_f-M_f^c|+const.\right)$.}\label{fig:hallkitaev}
\end{figure}
Fig.~\ref{fig:hallkitaev} shows the Hall conductance and its derivative
obtained by numerical integration of Eq.~(\ref{eq:halltwoband}) for a quench
starting from different $M_i$. As expected, if no quench is performed,
$C_{neq}$ equals the Chern number of the initial state. When $M_f$ approaches the
gap closing point $M_f^c = 0$, the curve becomes infinitely steep,
before it flattens again for smaller $M_f$. The derivative $dC_{neq}/dM_f$
is displayed together with fits of the form $\left(-|M_i-M_i^c|^{-1} \ln |M_f-M_f^c| + const.\right)$.
The numerical results form a perfect line as a function of $\ln |M_f-M_f^c |$
and thereby confirm the validity of above considerations regarding the
asymptotic behavior of $C_{neq}(M_f)$ nearby $M_f^c$.

Let us compare the nonanalytic behavior of the quench-state Hall conductance
in the Dirac model, the Haldane model, and the Kitaev model.
Despite of the significant difference in the symmetries and dispersion
relations of the models, the Hall conductance is always a continuous
function of $M_f$ and its derivative displays the following asymptotic behavior
\begin{equation}\label{eq:universalhallguess}
 \lim_{M_f \to M_f^c}\frac{d C_{neq}}{d M_f} \sim -\frac{K}{2}\frac{\ln|M_f-M_f^c|}{|M_i-M_i^c|},
\end{equation}
where $K$ is a constant. We already know $K=1$ in the Dirac model,
$K=-1$ in the Haldane model, and $K=2$ in the Kitaev model.
In fact, it is easy to see
that $K$ equals the change of the Chern number for the
ground state of the post-quench Hamiltonian at $M_f=M_f^c$.
This observation can be expressed as
\begin{equation}\label{eq:universalK}
 K = \lim_{\delta \to 0^+} C_f(M_f^c+\delta )-C_f(M_f^c-\delta),
\end{equation}
where $C_f$ denotes the Chern number for the ground state of $\hat H_f$.
In next section, we will strictly prove that
Eq.~(\ref{eq:universalhallguess}) and (\ref{eq:universalK})
stand for an arbitrary two-band Chern insulator.

\section{Universal nonanalytic behavior
of the Hall conductance in two-band Chern insulators}
\label{sec:generalnonanalytic}

Let us consider a two-band Chern insulator with
the single-particle Hamiltonian generally expressed as $\mathcal{H}_{\vec{k}}
=\vec{d}_{\vec{k}}\cdot \vec{\sigma}$. Recall that $\mathcal{H}_{\vec{k}}$
has two eigenvalues $\pm d_{\vec{k}}$ with $d_{\vec{k}}$ denoting
the length of the coefficient vector $\vec{d}_{\vec{k}}=\left({d}_{1\vec{k}},
{d}_{2\vec{k}},{d}_{3\vec{k}}\right)$.

Before discussing the nonanalyticity of the Hall conductance, we revisit the Chern number
for the ground state, which is expressed by using the coefficient vector as
\begin{equation}
 C= \int d\vec{k}^2 \frac{\left( \displaystyle\frac{\partial \vec{d}_{\vec{k}}}{\partial k_x}\times
 \frac{\partial \vec{d}_{\vec{k}}}{\partial k_y}\right) \cdot \vec{d}_{\vec{k}}}
 {4\pi d_{\vec{k}}^3}.
\end{equation}
One can easily see the topological nature of the Chern number
by reexpressing it as
\begin{equation}
 C = \frac{-1}{2\pi} \int d\vec{S} \cdot \left( \triangledown_{\vec{k}} \times \vec{A} \right),
\end{equation}
where
\begin{equation}\label{eq:berryconnectione}
 \vec{A}= \frac{d_{1\vec{k}}\triangledown_{\vec{k}} d_{2\vec{k}}
 -d_{2\vec{k}} \triangledown_{\vec{k}} d_{1\vec{k}}}
 {2d_{\vec{k}} (d_{\vec{k}}-d_{3\vec{k}})}
\end{equation}
is the so-called Berry connection and $\vec{S}$ denotes the Brillouin
zone oriented in the direction perpendicular to the momentum plane.
According to the Kelvin-Stokes theorem, the Chern number
equals the line integral of $\vec{A}$ along the
boundary of the Brillouin zone, plus the line integrals of $\vec{A}$
around all the singularities of $\vec{A}$ within the Brillouin zone.
The former integral must be zero due to the periodicity of $\vec{A}$.
Supposing that the singularities of $\vec{A}$ are
$\vec{q}_1, \vec{q}_2, \cdots, \vec{q}_N$, we then have
\begin{equation}\label{eq:cherngesum}
 C=\sum_{j=1}^N C^{(\vec{q}_j)}
\end{equation}
with
\begin{equation}\label{eq:chernnumberA}
C^{(\vec{q}_j)}=\frac{1}{2\pi} \lim_{\eta\to 0} \oint_{\partial B_\eta(\vec{q}_j)} \vec{A}\cdot d\vec{k},
\end{equation}
where $\partial B_\eta(\vec{q}_j)$ denotes the boundary of
a circle of radius $\sqrt{\eta}$ centered at $\vec{q}_j$, and the
integral is along the anticlockwise direction.

According to Eq.~(\ref{eq:berryconnectione}), a singularity of
$\vec{A}$ is a momentum $\vec{q}$ satisfying $d_{3\vec{q}}=d_{\vec{q}}$.
Note that $d_{\vec{k}}$ in the expression of $\vec{A}$
cannot be zero in a gapped state.
We then have $d_{1\vec{q}}=d_{2\vec{q}}=0$ at the singularity.
In fact, the tree components
of the coefficient vector are on an equal footing in the expression of
the Chern number. One can permute the three components
in the expression of $\vec{A}$. Therefore,
a singularity of $\vec{A}$ in general refers to a momentum at which
two components of the coefficient vector vanish. Here we
choose $d_{1\vec{q}}$ and $d_{2\vec{q}}$ as the vanishing components
without loss of generality.
For convenience of discussion, we call
$\vec{q}$ a singularity of $\vec{A}$ whether $d_{3\vec{q}}=d_{\vec{q}}$
or $d_{3\vec{q}}=-d_{\vec{q}}$. $d_{3\vec{q}}$ is a free parameter of
the model. We rename it as $m$ in next.
In fact, $m$ is nothing but $\left(M-M_c\right)$
in the Dirac model, the Haldane model or the Kitaev model in Sec.~\ref{sec:nonanalytic}.
Since $m=0$ indicates the closing of the energy gap, we
call $m$ the gap parameter.

In the case of multiple singularities in one Brillouin zone,
$m$ at different singularities might refer to different parameters
in the model. An example is the Haldane model, where a single Brillouin zone
contains two singularities $\vec{q}_1$ and $\vec{q}_2$ with
$d_{\vec{q}_1}\neq d_{\vec{q}_2}$. In a generic model,
$d_{\vec{k}}$ must have a global minimum point at one of the singularities,
when the system is close to the gap closing point.
In other words, the energy gap must be $2|m|$ at one of the singularities.
In fact, the minimum point of $d_{\vec{k}}$ is always related to the
symmetry of the model, so are the singularities.
On the other hand, if the energy gap closes simultaneously
at multiple singularities due to the symmetry of the model,
the gap parameter at these singularities must be the same one.
In this case, we say that
$m$ is the corresponding gap parameter for these singularities.
An example is the Kitaev honeycomb model in a magnetic field.

According to Eq.~(\ref{eq:cherngesum}) and (\ref{eq:chernnumberA}),
the Chern number is determined only by the coefficient vector
in the infinitesimal neighborhoods of the singularities.
Therefore, at each singularity $\vec{q}$, we expand
$\vec{d}_{\vec{k}}$ into a power series. Without loss of generality,
we have
\begin{equation}\label{eq:expdvector}
\begin{split}
d_{1\vec{k}} = & a_{1x} \Delta k_x + a_{1y} \Delta k_y + \mathcal{O}(\Delta k^2), \\
d_{2\vec{k}} = & a_{2x} \Delta k_x + a_{2y} \Delta k_y + \mathcal{O}(\Delta k^2), \\
d_{3\vec{k}} = & m + \mathcal{O}(\Delta k^2),
\end{split}
\end{equation}
where $\Delta \vec{k} := \vec{k}-\vec{q}$ and $a_{jx}$ and $a_{jy}$
depend on the singularity and the model.
A linear term is absent in the expression of $d_{3\vec{k}}$.
Otherwise, the minimum point of $d_{\vec{k}}$ would not be at
$\Delta \vec{k}=0$, which contradicts our assumption.
It is straightforward to verify the absence of the linear term in $d_{3\vec{k}}$
for the Dirac model, the Haldane model or the Kitaev model.

The Berry connection can be reexpressed as
\begin{equation}\label{eq:vecAexpexp}
 \vec{A}= \left(\frac{d_{\vec{k}}+d_{3\vec{k}}}{2d_{\vec{k}} }\right)
 \left(\frac{d_{1\vec{k}}\triangledown_{\vec{k}} d_{2\vec{k}}-d_{2\vec{k}}
 \triangledown_{\vec{k}} d_{1\vec{k}}}
 {\left(d_{1\vec{k}}\right)^2+\left(d_{2\vec{k}}\right)^2}\right).
\end{equation}
$ C^{(\vec{q})}$ is an integral of $\vec{A}$ over
the boundary of an infinitesimal neighborhood of $\vec{q}$.
When calculating $ C^{(\vec{q})}$,
we can replace the term inside the first bracket of Eq.~(\ref{eq:vecAexpexp})
by its value at $\Delta \vec{k}=0$, i.e. $(1+\textbf{sgn}(m))/2$.
The higher-order terms in the expansion of $d_{3\vec{k}}$
have no contribution to $ C^{(\vec{q})}$.
Regarding numerator and denominator inside the second bracket of Eq.~(\ref{eq:vecAexpexp}),
the terms $\mathcal{O}(\Delta k^2)$ in $d_{1\vec{k}}$
or $d_{2\vec{k}}$ contribute to the numerator a correction $\mathcal{O}(\Delta k^2)$
and to the denominator a correction $\mathcal{O}(\Delta k^3)$,
which can both be neglected when calculating
$\oint_{\partial B_\eta(\vec{q}_j)} \vec{A}\cdot d\vec{k}$ in the limit $\eta \to 0$.
Therefore, $ C^{(\vec{q})}$ is only determined by the lowest order expansion
of the coefficient vector given by Eq.~(\ref{eq:expdvector}).
It is straightforward to determine $ C^{(\vec{q})}$ as
\begin{equation}\label{eq:cvecq}
 C^{(\vec{q})}= \frac{1}{2} \left( 1+\textbf{sgn}(m) \right) \textbf{sgn}
 \left( a_{1x}a_{2y}-a_{2x}a_{1y}\right).
\end{equation}
At the gap closing point $m=0$, the change of $C^{(\vec{q})}$
is $\textbf{sgn} \left( a_{1x}a_{2y}-a_{2x}a_{1y}\right)$.

Next we discuss the nonanalyticity of the quench-state Hall conductance,
which can be expressed by using the initial and post-quench coefficient vectors as
\begin{equation}
  C_{neq} = \int d \vec{k}^2 
  \frac{\left(\vec{d}^f_{\vec{k}} \cdot \vec{d}^i_{\vec{k}} 
  \right) \left( \displaystyle\frac{\partial \vec{d}^f_{\vec{k}}}{\partial k_x}\times
 \frac{\partial \vec{d}^f_{\vec{k}}}{\partial k_y}\right) \cdot \vec{d}^f_{\vec{k}}}
 {4\pi {d}^i_{\vec{k}} (d^f_{\vec{k}})^4}. 
\end{equation}
The integral in the expression of $C_{neq}$ is over the Brillouin zone
which is finite in general.
If the integrand is an analytic
function of $\vec{k}$ and the other parameters in the Hamiltonians
and is bounded in the Brillouin zone, $C_{neq}$ must be analytic.
While in a generic model, the coefficient vectors $\vec{d}^i_{\vec{k}}$
and $\vec{d}^f_{\vec{k}}$ are analytic functions and are bounded.
Therefore, $C_{neq}$ is nonanalytic only if $d^f_{\vec{k}}$ vanishes somewhere in
the Brillouin zone, that is the energy gap of the post-quench
Hamiltonian closes. Notice that, without loss of generality,
the initial state is chosen to be a gapped state so that ${d}^i_{\vec{k}}$
cannot be zero.

As mentioned above, the energy gap closes only at the singularities.
Without loss of generality, we suppose that $m$ is the corresponding
gap parameter for the singularities $\vec{q}_1, \vec{q}_2, \cdots,
\vec{q}_{N'}$ with $N'\leq N$. The Hall conductance $C_{neq}$
as a function of $m_f$ (the gap parameter for the post-quench Hamiltonian)
is nonanalytic at $m_f=0$. We divide the domain of integration in
the expression of $C_{neq}$ into
the infinitesimal neighborhoods of $\vec{q}_1, \vec{q}_2, \cdots,
\vec{q}_{N'}$ and the left area.
The nonanalytic part of the Hall conductance is expressed as
\begin{equation}\label{eq:sigmaheta}
C^\eta_{neq} = \sum_{j=1}^{N'} C_{neq}^{(\vec{q}_j)}
\end{equation}
with
\begin{equation}\label{eq:hallnonana}
\begin{split}
C_{neq}^{(\vec{q}_j)} = \int_{\mathcal B_\eta(\vec{q}_j)} d \vec{k}^2 
  \frac{\left(\vec{d}^f_{\vec{k}} \cdot \vec{d}^i_{\vec{k}} 
  \right) \left( \displaystyle\frac{\partial \vec{d}^f_{\vec{k}}}{\partial k_x}\times
 \frac{\partial \vec{d}^f_{\vec{k}}}{\partial k_y}\right) \cdot \vec{d}^f_{\vec{k}}}
 {4\pi {d}^i_{\vec{k}} (d^f_{\vec{k}})^4}, 
 \end{split}
\end{equation}
where $\mathcal B_\eta(\vec{q}_j)$ is a circle of radius $\sqrt{\eta}$ centered at
$\vec{q}_j$. $\eta$ can be arbitrarily small. While $\left(C_{neq}-C^\eta_{neq}\right)$
is an analytic function of $m_f$, since $d^f_{\vec{k}}$
has a positive lower limit in the corresponding domain
of integration.

$C_{neq}^{(\vec{q})}$ is an integral over the infinitesimal
neighborhood of $\vec{q}$. For calculating $C_{neq}^{(\vec{q})}$
we expand the coefficient vectors $\vec{d}^i_{\vec{k}}$ and $\vec{d}^f_{\vec{k}}$
around $\vec{q}$. Let us first consider the lowest order expansion given by
Eq.~(\ref{eq:expdvector}). ${d}^i_{\vec{k}}$ can be replaced by
its value at $\vec{k}=\vec{q}$, i.e. ${d}^i_{\vec{q}}=\left|m_i\right|$ with
$m_i$ denoting the gap parameter in the initial Hamiltonian.
The energy gap in the initial state is $2|m_i|$. Substituting
Eq.~(\ref{eq:expdvector}) into Eq.~(\ref{eq:hallnonana}),
for the denominator we obtain
\begin{equation}\label{eq:denoexp}
 \left( d_{\vec{k}}^f\right)^4 = \left( m_f^2 +
 \sum_{j=1}^2 \left( a_{jx} \Delta k_x + a_{jy}\Delta k_y \right)^2 \right)^2.
\end{equation}
We already assumed that $d_{\vec{k}}^f$ has an isolated minimum point
at the singularity $\Delta \vec{k}=0$. The spectrum must have a conic structure
nearby the singularity. For $a_{jx}$ and $a_{jy}$ that meet this constraint,
we can always perform a linear transformation of coordinates to get
\begin{equation}
 \sum_{j=1}^2 \left( a_{jx} \Delta k_x + a_{jy}\Delta k_y \right)^2 = \Delta k'^2.
\end{equation}
In the new coordinates, for the numerator of the integrand
in Eq.~(\ref{eq:hallnonana}) we obtain
\begin{equation}
 \begin{split}
nu. = & (a_{1x}a_{2y}-a_{2x}a_{1y}) \bigg[ m_im_f^2+m_f \Delta k'^2 \bigg].
 \end{split}
\end{equation}
It is then straightforward to work out $C_{neq}^{(\vec{q})}$, which is
\begin{equation}
 C_{neq}^{(\vec{q})} \sim \frac{-\textbf{sgn}(a_{1x}a_{2y}-a_{2x}a_{1y})}{2|m_i|}
 m_f \ln |m_f|.
\end{equation}
Here we only show the nonanalytic part of $C_{neq}^{(\vec{q})}$,
which is independent of the shape of the neighborhood or
the choice of $\eta$.

Clearly, $C_{neq}^{(\vec{q})}$ is a continuous function of $m_f$
even at $m_f=0$. The asymptotic behavior of its derivative
in the limit $m_f \to 0$ is
\begin{equation}\label{eq:cneqqtomf}
\lim_{m_f\to 0} \frac{dC_{neq}^{(\vec{q})}}{dm_f} \sim
 \frac{-\textbf{sgn}(a_{1x}a_{2y}-a_{2x}a_{1y})}{2|m_i|}
 \ln |m_f|.
\end{equation}
Recall that $\textbf{sgn}(a_{1x}a_{2y}-a_{2x}a_{1y})$
equals the change of $C^{(\vec{q})}$ at the gap closing point.
$C^{(\vec{q})}$ is the contribution
to the Chern number from the singularity $\vec{q}$.
For distinguishing $C^{(\vec{q})}$ for the ground state of
$\hat H_f$ from that for the ground state of $\hat H_i$,
the former is specifically denoted by the symbol $C^{(\vec{q})}_f$.
$C^{(\vec{q})}_f$ is given
by Eq.~(\ref{eq:cvecq}) in which $m$ is replaced by $m_f$.
We can then express Eq.~(\ref{eq:cneqqtomf}) by using
the change of $C^{(\vec{q})}_f$ at $m_f=0$ as
\begin{equation}
\lim_{m_f\to 0} \frac{dC_{neq}^{(\vec{q})}}{dm_f} \sim
 \frac{\displaystyle\lim_{m_f\to 0^-} C^{(\vec{q})}_f(m_f)
 -\lim_{m_f\to 0^+} C^{(\vec{q})}_f(m_f)}{2|m_i|}
 \ln |m_f|.
\end{equation}

The nonanalytic part of the Hall conductance at $m_f=0$ is a sum of
$C_{neq}^{(\vec{q})}$ at the singularities $\vec{q}_1, \vec{q}_2, \cdots,
\vec{q}_{N'}$ that corresponds to $m_f$.
While the Chern number $C_f$ is a sum of
$C^{(\vec{q})}_f$ at all the singularities $\vec{q}_1, \vec{q}_2, \cdots,
\vec{q}_{N}$ in the Brillouin zone.
The singularity $\vec{q}_j$ with $j>N'$ does not correspond to $m_f$,
and then has no contribution to the nonanalyticity of $C_{neq}$.
But $C^{(\vec{q}_j)}_f$ also keeps invariant at $m_f=0$,
since $C^{(\vec{q}_j)}_f$ changes only if its corresponding gap parameter
becomes zero.
Therefore, the asymptotic behavior
in the derivative of the Hall conductance can be expressed as
\begin{equation}\label{eq:central}
 \lim_{m_f\to 0} \frac{dC_{neq}}{dm_f} \sim
 \frac{\displaystyle\lim_{m_f\to 0^-} C_f(m_f)- \displaystyle\lim_{m_f\to 0^+} C_f(m_f) }{2|m_i|}
 \ln |m_f|.
\end{equation}
Eq.~(\ref{eq:central}) is equivalent to Eq.~(\ref{eq:universalhallguess}).
Note that the gap parameter $m_{i/f}$ is just $\left(M_{i/f}-M^c_{i/f}\right)$ defined
in the Dirac model, the Haldane model, and the Kitaev model.
We thus prove that the nonanalytic behavior of the Hall conductance
found in these three models is universal in a generic two-band Chern insulator.
The quench-state Hall conductance is a continuous function of
the gap parameter in the post-quench Hamiltonian.
Its derivative is logarithmically divergent as the gap parameter becomes zero.
The prefactor of the logarithm
is the ratio of the change of the Chern number for the ground state
of the post-quench Hamiltonian to the energy gap in the initial state.
The Hall conductance is nonanalytic only at the
gap closing points where the Chern number changes.

Up to now, we only considered the lowest order expansion of the
coefficient vector in the calculation of $C^{(\vec{q})}_{neq}$.
To finish our proof, we will show that the
higher-order terms in the expansion do not affect the continuity
of $C^{(\vec{q})}_{neq}$ and have no contribution to
the asymptotic behavior of $dC^{(\vec{q})}_{neq}/dm_f$ in the limit $m_f \to 0$.

Let us consider the second order term in the expansion of $d_{3\vec{k}}$
(see Eq.~(\ref{eq:expdvector})).
Without loss of generality, we suppose it to be
$\Omega=b_{3x}\Delta k_x^2 + b_{3y}\Delta k_y^2 + b_{3m} \Delta k_x \Delta k_y $
with $b_{3x}, b_{3y}$ and $b_{3m}$ denoting the free parameters.
We recalculate the integrand in the expression of $C_{neq}^{(\vec{q})}$
(see Eq.~(\ref{eq:hallnonana})). For the denominator we obtain
\begin{equation}\label{eq:cneqdenohigh}
\begin{split}
 \left( d_{\vec{k}}^f\right)^4 = & \bigg[ m_f^2 +
 \sum_{j=1}^2 \left( a_{jx} \Delta k_x + a_{jy}\Delta k_y \right)^2 \\ &
 + 2m_f\left(b_{3x}\Delta k_x^2 + b_{3y}\Delta k_y^2 + b_{3m} \Delta k_x \Delta k_y \right)
\\ &  + \mathcal{O}(\Delta k^4) \bigg]^2.
 \end{split}
\end{equation}
For the numerator we obtain
\begin{equation}
 \begin{split}
nu.
 = & (a_{1x}a_{2y}-a_{2x}a_{1y}) \\ & \times \bigg[ m_im_f^2 +m_f \sum_{j=1}^2 \left(a_{jx}\Delta k_x
 +a_{jy}\Delta k_y\right)^2 \\ & + m_f^2 \left(b_{3x}\Delta k_x^2+
b_{3y}\Delta k_y^2+ b_{3m}\Delta k_x \Delta k_y\right) + \mathcal{O}(\Delta k^4) \bigg].
 \end{split}
\end{equation}
Since the integral
is over an infinitesimal neighborhood of $\vec{q}$,
the fourth order terms $\mathcal{O}(\Delta k^4)$ in numerator
and denominator of the integrand can be neglected.
The additional second order term in $ \left( d_{\vec{k}}^f\right)^4$
that comes from $\Omega$ is proportional to $m_f$.
Therefore, it is much smaller than the other second order terms
in the limit $m_f\to 0$ and can be neglected in the study of
the nonanalyticity at $m_f=0$.
For the numerator, the additional second order term
coming from $\Omega$ is proportional to $m_f^2$ and then
can be neglected, since the other second
order terms in the numerator is proportional to $m_f$.
A more precise argument can be obtained by replacing
$\sum_{j=1}^2 \left( a_{jx} \Delta k_x + a_{jy}\Delta k_y \right)^2$
by $\Delta k'^2$ and replacing $\Omega$ by $b \Delta k'^2$ both in numerator and denominator
of the integrand.
After this replacement, the integral can be worked out.
It is then straightforward to verify that $\Omega$
does not change the continuity of $C_{neq}^{(\vec{q})}$
or the asymptotic behavior of $dC_{neq}^{(\vec{q})}/dm_f$.
Furthermore, in the power series of $d_{3\vec{k}}$, the terms of order
higher than two contribute to numerator or denominator of the integrand
the corrections which are at least of order three. These
corrections can be neglected for an infinitesimal domain
of integration.

Next we consider the higher-order terms in $d_{1\vec{k}}$
or $d_{2\vec{k}}$. The terms of order higher than
one contribute to the denominator of the integrand
a correction $\mathcal{O}(\Delta k^3)$.
At the same time, the terms of order higher than three
contribute to the numerator a correction $\mathcal{O}(\Delta k^3)$.
Therefore, the terms of order higher than three in $d_{1\vec{k}}$
or $d_{2\vec{k}}$ can be neglected.
The second and third order terms in $d_{1\vec{k}}$
or $d_{2\vec{k}}$ contribute to the numerator a linear term and a second order term that
is proportional to $m_f^2$.
The latter can be neglected due to the same reason mentioned above.
Finally, the linear term in the numerator is antisymmetric
under $\Delta k_\alpha \to -\Delta k_\alpha$.
The integral of an antisymmetric function over a circle
must be zero. Hence, the terms of order higher than one
in $d_{1\vec{k}}$ or $d_{2\vec{k}}$ can be neglected.
Thereby, we finished our proof.
Both the Chern number for the ground state and
the nonanalyticity of the Hall conductance for the quenched state
are only determined by the lowest order expansion of
the coefficient vector at the singularities.

Finally, it is worth emphasizing that
the asymptotic behavior of $dC_{neq}/dm_f$
depends only upon the change of the Chern number at $m_f=0$ and
the energy gap in the initial state.
The universal nonanalytic behavior of the Hall conductance
is related to the conic structure of the spectrum
at the singularities.
It is topologically protected in the sense that
it is independent of the detail of the model.
Therefore, even if we only consider the
noninteracting model in the absence of disorder in the above argument,
the nonanalytic behavior of the Hall conductance should not be
changed by weak interaction or weak disorder.
But the Chern number in Eq.~(\ref{eq:central}) is not
well-defined in the presence of the interaction between particles.
It is then necessary to find a more generic topological invariant instead of $C_f$.
This generic topological invariant is argued to be the winding number of the Green's function
in next section.

\section{Topological invariant for the quenched state}
\label{sec:topologicalinvariant}

Now we are prepared to discuss which is the experimentally-relevant
topological invariant for the quenched state of a Chern insulator. A naive idea
of defining the topological invariant
for a quenched state is to use the Chern number of the
unitarily-evolving wave function. It is defined as
\begin{equation}\label{realtimechern}
 C(t) = \frac{i}{2\pi} \sum_{\alpha \in oc} \int d\vec{k}^2 \left( \Braket{ \frac{\partial 
 u_{\vec{k}\alpha}(t)}{\partial k_x} | \frac{\partial u_{\vec{k}
 \alpha}(t)}{\partial k_y}} - \text{H.c.}\right).
\end{equation}
$C(t)$ is a straightforward extension of the Chern number for the
ground state given by
Eq.~(\ref{eq:cherngeneral}) in which the eigenstate $\ket{u_{\vec{k}
 \alpha}}$ is replaced by the evolving single-particle state $\ket{u_{\vec{k}
 \alpha}(t)}$. The sum of $\alpha$ is over the occupied bands
in the initial state. $C(t)$ is well-defined for a noninteracting
model in which the evolution of different single-particle states
is independent of each other. But $C(t)$
keeps invariant after a quench, being independent of
the post-quench Hamiltonian $\hat H_f$~\cite{alessiol,caio,wang15}.
Therefore, one cannot use $C(t)$ to explain the nonequilibrium
phase transition in the quenched states, which is indicated by the nonanalyticity of the Hall conductance
as $\hat H_f$ changes.

It has been shown that the Chern number $C_f$ for the
ground state of the post-quench Hamiltonian can be used as the topological
invariant for the quenched state. It is experimentally relevant
in the sense that the nonequilibrium
phase transition is always accompanied by the change of $C_f$
and vice versa. The change of $C_f$
also determines the asymptotic behavior of the derivative
of the Hall conductance at the transition. But $C_f$ is not
well-defined in the presence of interaction.
On the other hand, it is well known that
the winding number of the Green's function $W$ is equal to
the Chern number for the ground states~\cite{niu}. Next we will
show that, for the quenched states in a noninteracting model,
$W$ in fact equals the Chern number $C_f$, being independent
of the initial state. Since $W$ is also well-defined
in the presence of interaction, it serves as a more generic
topological invariant for the quenched state.
We notice that $W$ was already employed to describe
the topological property of the quenched state in a topological
superfluid~\cite{fosterprb}.

The textbook definition of the retarded Green's function after a quench is
\begin{equation}\label{eq:retarededtime}
 G^r_{j,j'}(\vec{k},t,t')=- i 
 \theta(t-t') \bra{\Psi(0)} \{ \hat c_{\vec{k}j}(t)
 , \hat c^\dag_{\vec{k}j'}(t') \}_+\ket{\Psi(0)},
\end{equation}
where $\ket{\Psi(0)}$ denotes the initial state,
$\hat c_{\vec{k}j}$ with $j=1,2,\cdots,N$ denotes
the fermionic operator in the original basis of the Hamiltonian~(\ref{eq:orginaloperatorH}),
and $\{\}_+$ denotes the anticommutator.
The Green's function in frequency-momentum space is obtained by a Fourier transformation as
\begin{equation}
 G^r_{j,j'}(\omega,\vec{k})= \int_{-\infty}^\infty d(t-t')
  e^{i\omega (t-t')} G^r_{j,j'}(\vec{k},t,t').
\end{equation}
Note that $t,t'>0$ must be larger than the time when the quench is performed.
The domain of the Green's function can be extended to imaginary frequency by
analytic continuation. Let us use $G^r$ to denote
the $N$-by-$N$ Green's function matrix with the elements $ G^r_{j,j'}(i\omega,\vec{k})$.
The winding number is then defined as~\cite{niu,gurarie11}
\begin{equation}
\begin{split}
 W=&\frac{1}{24\pi^2} \int_{-\infty}^{\infty} d\omega
 \int d\vec{k}^2 \\ & \times \epsilon^{\alpha\beta\gamma}
 \textbf{Tr}\left[ G^{r-1} \frac{\partial G^r}{\partial k_\alpha}
G^{r-1} \frac{\partial G^r}{\partial k_\beta}
G^{r-1} \frac{\partial G^r}{\partial k_\gamma}\right],
\end{split}
\end{equation}
where $\epsilon^{\alpha\beta\gamma}$ is the Levi-Civita symbol with
$\alpha,\beta,\gamma=0,1,2$, and $k_0=\omega$, $k_1=k_x$ and $k_2=k_y$.
$G^{r-1}$ is the inverse of $G^r$.

In Eq.~(\ref{eq:retarededtime}), the time-dependent operators are defined as
$\hat c_{\vec{k}j}(t)=e^{i\hat H_f t} \hat c_{\vec{k}j} e^{-i\hat H_f t}$.
In the absence of interaction, the post-quench Hamiltonian $\hat H_f$ is quadratic.
$\hat c_{\vec{k}j}(t)$ must be a linear combination of $\hat c_{\vec{k}j'}$
with $j'=1,2,\cdots,N$. Therefore, $\{ \hat c_{\vec{k}j}(t)
 , \hat c^\dag_{\vec{k}j'}(t') \}_+$ is in fact a number instead of an operator.
The Green's function is then independent of the initial state $\ket{\Psi(0)}$,
so is the winding number $W$. $W$
depends only upon the post-quench Hamiltonian $\hat H_f$.
$W$ keeps invariant even if we replace $\ket{\Psi(0)}$ by the ground
state of $\hat H_f$. Therefore,
$W$ must equal the Chern number $C_f$ for the ground state of $\hat H_f$~\cite{niu}.

We reexpress the asymptotic behavior of the derivative of the Hall conductance
by using $W$ as
\begin{equation}\label{eq:centralW}
 \lim_{m_f\to 0} \frac{dC_{neq}}{dm_f} \sim
 \frac{\displaystyle\lim_{m_f\to 0^-} W(m_f)- \displaystyle\lim_{m_f\to 0^+} W(m_f) }{2|m_i|}
 \ln |m_f|.
\end{equation}
We expect that Eq.~(\ref{eq:centralW}) also stands in the presence of
weak interaction. Moreover, $2\left|m_i\right|$ and $2\left|m_f\right|$
in the formula denote the energy gap of $\hat H_i$ and $\hat H_f$,
respectively, which are well-defined even in the presence of interaction.
The change of the winding number determines the
nonanalyticity of the Hall conductance at the gap closing point of $\hat H_f$.
Whether there is a nonequilibrium phase transition in the quenched state is
uniquely determined by whether the winding number changes.
In this sense, the nonequilibrium phase transition is topologically driven.
And the winding number is the experimentally-relevant
topological invariant for the quenched state.

\section{Conclusions}
\label{sec:con}

In summary, the Hall conductance of a quenched state in the long time limit
is calculated by applying the linear response theory to the diagonal ensemble.
In the eigenbasis of the post-quench Hamiltonian,
the diagonal ensemble is obtained by neglecting all the off-diagonal
elements in the initial density matrix.
The quench-state Hall conductance can be expressed as the integral
of the Berry curvature weighted by the nonequilibrium distribution of particles
(see Eq.~(\ref{eq:reducedHall})).
It is not quantized in general, but can take an arbitrary value.

The Hall conductance as a function of the gap parameter $m_f$
in the post-quench Hamiltonian $\hat H_f$ displays a universal
nonanalytic behavior in a generic two-band Chern insulator.
The examples include
the Dirac model, the Haldane model,
and the Kitaev honeycomb model in the fermionic basis.
The Hall conductance is continuous everywhere.
But its derivative with respect to $m_f$
is logarithmically divergent in the limit $m_f\to 0$
(the energy gap of $\hat H_f$ is $2\left|m_f\right|$),
if the winding number of the Green's function for the
quenched state $W$ changes at $m_f=0$.
The prefactor of the logarithm is the ratio of the
change of $W$ to the energy gap in the initial state (see Eq.~(\ref{eq:centralW})).
The nonanalyticity of the Hall conductance indicates
a topologically driven nonequilibrium phase transition.
The topological invariant for the quenched state
is the winding number $W$.

The Hamiltonian of a two-band Chern insulator in momentum space can
be decomposed into the linear combination of Pauli matrices.
The nonanalyticity of the Hall conductance depends only upon
the lowest order expansion of the coefficients of Pauli matrices
at the singularities in the Brillouin zone.
Singularities are defined as momenta where the energy gap closes at $m_f=0$.

The Haldane model has been realized
in an optical lattice recently~\cite{jotzu}.
A system of cold atoms is well isolated from the environment.
The nonequilibrium distribution of atoms in a quenched state
survives for a long time.
Our prediction can therefore be checked in a system of cold atoms.
It is difficult to measure the Hall conductance directly
in cold atoms. But the Hall conductance can also be obtained from
the Faraday rotation angle~\cite{qi10}, which is correspondingly easier to measure in cold atoms.
On the other hand,
in solid-state materials where the measurement of Hall conductance is a
standard technique, the nonequilibrium
distribution of electrons is difficult to realize due to the
fast relaxation process. A possible solution is to
periodically drive the system for keeping it
out of equilibrium. The time evolution of
a periodically-driven system is governed by
a time-independent Floquet Hamiltonian~\cite{kitagawa}.
It was suggested that a Floquet Chern insulator can be
realized in graphene ribbons~\cite{oka08,gu11}.
Due to the similarity between the dynamics of a periodically-driven quantum state
and a quenched state~\cite{lazarides}, we expect that
the techniques developed in this paper can also be used
to analyze the nonanalyticity of the Hall conductance
in a Floquet Chern insulator.

\section*{Acknowledgement}
This work is supported by NSFC under Grant No.~11304280 and China Scholarship Council.
M.S. acknowledges support by the German National Academic Foundation.

\end{document}